\documentclass[prb,twocolumn,floatfix,showpacs,amsmath,amssymb,superscriptaddress]{revtex4-1}

\usepackage{graphicx}
\usepackage{epstopdf}
\usepackage{dcolumn}
\usepackage{bm}
\usepackage{hyperref}

\newcommand{\ket}[1]{\begingroup\lvert#1\rangle\endgroup}

\begin{document}

\title{Supercurrent and Andreev bound state dynamics in superconducting
quantum point contacts under microwave irradiation}

\author{F.S. Bergeret$^{1,2}$,P. Virtanen$^{3,4}$, A. Ozaeta$^{1}$, T.T. Heikkil\"{a}$^{4}$, J.C. Cuevas$^5$}

\affiliation{
Centro de F\'{\i}sica de Materiales (CFM-MPC), Centro Mixto
CSIC-UPV/EHU, Manuel de Lardizbal 5, E-20018 San Sebasti\'an, Spain\\
$^{2}$Donostia International Physics Center (DIPC), Manuel de Lardizbal 4,
E-20018 San Sebasti\'an, Spain\\
$^{3}$Institute for Theoretical Physics and Astrophysics,
University of W\"urzburg, D-97074 W\"urzburg, Germany\\
$^4$Low Temperature Laboratory, Aalto University, 
P.O. Box 15100, FI-00076 AALTO, Finland\\
$^5$Departamento de F\'{\i}sica Te\'orica de la Materia Condensada,
Universidad Aut\'onoma de Madrid, E-28049 Madrid, Spain}

\date{\today}

\begin{abstract}
We present here an extensive theoretical analysis of the supercurrent of a
superconducting point contact of arbitrary transparency in the presence of
a microwave field. Our study is mainly based on two different approaches: a two-level
model that describes the dynamics of the Andreev bound states in these systems
and a fully microscopic method based on the Keldysh-Green function technique.
This combination provides both a deep insight into the physics of irradiated
Josephson junctions and quantitative predictions for arbitrary range of parameters.
The main predictions of our analysis are: (i) for weak fields and low temperatures,
the microwaves can induce transitions between the Andreev states leading to a large
suppression of the supercurrent at certain values of the phase, (ii) at strong
fields, the current-phase relation is strongly distorted and the corresponding
critical current does not follow a simple Bessel-function-like behavior, and (iii)
at finite temperature, the microwave field can enhance the critical current
by means of transitions connecting the continuum of states outside the gap region
and the Andreev states inside the gap. Our study is of relevance for a large
variety of superconducting weak links as well as for the proposals of using
the Andreev bound states of a point contact for quantum computing applications.
\end{abstract}


\maketitle

\section{Introduction}

In 1962 Josephson predicted that a dissipationless current (supercurrent)
could flow in a junction between two superconductors (S) weakly coupled by an
insulating barrier,\cite{Josephson1962} which was confirmed experimentally
shortly afterwards by Anderson and Rowell.\cite{Anderson1963} Soon after this
confirmation, it became clear that this phenomenon, referred to as the dc Josephson
effect, could take place in a variety of superconducting weak links such as Dayem
bridges, SNS junctions, where N corresponds to a normal metal bridge, or large
point contacts.\cite{Likharev1979,Barone1982} The only difference between these
systems lies on the exact current phase relation (CPR), which depends on
the characteristics of the constriction linking the superconducting
leads.\cite{Golubov2004}

In recent years, the dc Josephson effect has been investigated in novel
superconducting junctions with weak links based on atomic
contacts,\cite{Koops1996,Goffman2000,Rocca2007} carbon
nanotubes,\cite{Kasumov1999,Jarillo-Herrero2006,Jorgensen2006}
fullerenes,\cite{Kasumov2005} semiconductor nanowires,\cite{Doh2005,Xiang2006} or
graphene.\cite{Heersche2007,Du2008,Jeong2011} Some of these nanostructures fall into
the category of a superconducting quantum point contact (SQPC), where the constriction
has a length much smaller than the superconducting coherence length. In this limit,
and in the absence of strong interactions in the constriction, the dc Josephson
effect can be described in a unified manner using two basic concepts of mesoscopic
physics, namely the concepts of conduction channels and Andreev bound states. In the
normal state, the coherent transport through a mesoscopic system can be described
in terms of the independent contributions of the eigenfunctions of the transmission
matrix from the structure, known as conduction channels, and these contributions are determined
by the corresponding transmission coefficients $\{\tau_i\}$. In the superconducting
state, the electrons (holes) transmitted in a conduction channel are Andreev
reflected at the electrodes as holes (electrons) in the same channel. This process
is successively repeated in both electrodes leading to the formation of a pair of
bound states in the gap region. These are known as the Andreev bound states (ABSs). In the case
of a single-channel SQPC with transmission $\tau$, the energies of the ABSs are given
by\cite{Furusaki1991,Beenakker1991}
\begin{equation}
\label{eq-ABS}
E^{\pm}_{\rm A}(\varphi,\tau) = \pm E_{\rm A}(\varphi,\tau) =
\pm \Delta \sqrt{1 - \tau \sin^2 (\varphi/2)} ,
\end{equation}
where $\Delta$ is the superconducting gap and $\varphi$ is the phase difference
between the order parameters on both sides of the junction. In equilibrium,
these two states carry opposite supercurrents $I^{\pm}_{\rm A} (\varphi) =
(2e/\hbar) \partial E^{\pm}_{\rm A}/ \partial \varphi$, which are weighted by
the occupation of the ABSs (determined by the Fermi function). In the case of
a multichannel SQPC, the supercurrent is simply given by the sum of the contributions
from the individual channels.\cite{Beenakker1991}

This unified microscopic picture of the dc Josephson effect has been
confirmed experimentally in the context of atomic contacts by Della Rocca
and coworkers.\cite{Rocca2007} In particular, these authors measured the CPR of
an atomic contact placed along  with a tunnel junction in a small superconducting loop
and found an excellent agreement with the theory using the independently determined
transmission coefficients. At this stage, one may wonder whether it is possible
to control the occupation of ABSs of a SQPC with an external field, and in turn to
control the supercurrent. This is the main issue  explored in this
work and for this purpose, we present here an extensive theoretical analysis
of the supercurrent and the dynamics of the ABSs of a SQPC under microwave irradiation.
This is a basic problem in mesoscopic superconductivity, which is also relevant
for the field of quantum computing since the ABSs of a SQPC have been proposed to be used as the two states of a  qubit.\cite{Desposito2001,Zazunov2003,Zazunov2005} In this proposal,
a microwave field can be used for the spectroscopy of the two-level system or
to probe its quantum state by current measurements.

The microwave-assisted supercurrent in SQPCs is often discussed in the framework
of the adiabatic approximation (see Section II), where one assumes that the ABSs
follow adiabatically the microwave field. This approximation does not take into
account the possible transitions between the ABSs and therefore, if fails
to describe the current at high frequencies or for highly transmissive
contacts, where the energy difference  between the states can be rather small. The first
microscopic analysis of this problem for a SQPC of an arbitrary transparency was
reported by Shumeiko and coworkers.\cite{Shumeiko1993} These authors studied the
limit of weak fields and predicted the possibility to have a large suppression
of the current due to resonant transitions between the ABSs. Later, other
aspects of this problem, including the dynamics of the ABSs, have been addressed
focusing on the linear response regime.\cite{Gorelik1995,Gorelik1998,Lundin2000}
A complete solution of this problem, valid for an arbitrary range of parameters,
has only been reported very recently.\cite{Bergeret2010} In this latter work, we
developed a theory of the supercurrent through a microwave-irradiated SQPC in
the framework of the Keldysh-Green function technique. This theory allowed us to
put forward new predictions such as the evolution of the CPR with the radiation power
and  the possibility to enhance the critical current at finite temperatures
by irradiating the junction. Here, we  describe in detail this theory (see
Section IV) and, in particular, we  present new analytical results that elucidate
the origin of the microwave-enhanced supercurrents.

In the process of understanding the results of the exact theory, we are confronted
with the question of to what extent the physics of microwave-irradiated SQPCs can
be understood in terms of just the dynamics of the ABSs, i.e., in terms of a
natural extension of the argument described in the previous paragraphs for the
case of a junction in equilibrium. In order to answer this question, we 
make use of the two-level Hamiltonians of a SQPC existing in the
literature\cite{Ivanov1999,Zazunov2003} and we  compare the results with the
exact theory. This comparison serves in turn to establish the range of validity
of these two-level models. It is worth stressing that within these
models the computation of the dc properties such as the supercurrent or the average occupation of the ABSs
for arbitrary radiation is a highly non-trivial task. In order to carry it out,
we have developed a new powerful method which allows us to compute any dc
quantity in an arbitrary two-level system driven out of equilibrium by a periodic
perturbation. This method is described in Section III and it constitutes one of
the main results of this work. With the help of this method, we show
that with the Hamiltonian of Ref.~\onlinecite{Zazunov2003} one can nicely reproduce the
exact results at low temperatures and low radiation powers. Moreover, this analysis
allows us to obtain analytical results for the supercurrent dips produced by
microwave-induced transitions between the ABSs.

The rest of the paper is organized as follows. In the next section we  briefly
review the equilibrium properties of a SQPC as well as the basic results of the adiabatic
approximation. In Section III we  study the dynamics of the ABSs under a microwave
field within the two-level Hamiltonian of Ref.~\onlinecite{Zazunov2003}. In particular,
we  describe a novel method that allows us to obtain the CPR for any power and
frequency of the external field, and we  also derive analytical expressions for
the supercurrent beyond the rotating wave approximation. In Section IV we  discuss
the Keldysh-Green function technique, which describes the supercurrent for an arbitrary
range of parameters, including also the contribution of the continuum of states outside
the gap region. We  present a detailed comparison of the results of this
technique at zero temperature with those obtained with the two-level model. Moreover,
we  analyze in detail the phenomenon of microwave-enhanced supercurrent at
finite temperatures, for which we  present analytical results. Finally, Section 
V is  devoted to some additional discussions and to summarizing the main results of
this work.

\section{System and adiabatic approximation}\label{adiabtic}

We consider a SQPC consisting of two identical superconducting electrodes with
an energy gap $\Delta$,  linked by a single conduction channel of transmission
$\tau$. Our main goal is to compute the supercurrent through this system when it
is subjected to a monochromatic microwave field of frequency $\omega$.  We assume that the external radiation generates a time-dependent voltage $V(t) =
V_0 \sin \omega t$,\cite{Barone1982} where the amplitude $V_0$  depends on the
power of the external radiation source, and eventually also on the polarization
of the radiation. According to the Josephson relation, this voltage induces a
time-dependent superconducting phase difference given by
\begin{equation}
\phi(t) = \varphi + 2 \alpha \cos \omega t , \label{eq-phi}
\end{equation}
where $\varphi$ is the dc part of the phase and $\alpha = eV_0/\hbar \omega$ is
a parameter that measures the strength of the coupling to the electromagnetic
field and it is used  here as a parameter to be determined by comparing with
the experiments.

As explained in the introduction, in the absence of microwaves the supercurrent
can be expressed as a sum of the contributions of the two ABSs as $I_{\rm eq}(\varphi)
= I^-_{\rm A} n_{\rm F}(E^-_{\rm A}) + I^+_{\rm A} n_{\rm F}(E^+_{\rm A})$,
$n_{\rm F}(E)$ being the Fermi distribution function, which yields\cite{Haberkorn1978}
\begin{equation}
\label{Ieq}
I_{\rm eq}(\varphi) = \frac{e\Delta^2}{2\hbar} \frac{\tau \sin\varphi}
{E_{\rm A}(\varphi)} \tanh \left( \frac{E_{\rm A}(\varphi)} {2k_{\rm B}T} \right) ,
\end{equation}
where $E_{\rm A}$ is defined in Eq.~(\ref{eq-ABS}) and $T$ is the temperature.
In the tunnel regime ($\tau \ll 1$), this expression reduces to the sinusoidal CPR
given by the Ambegaokar-Baratoff formula:\cite{Ambegaokar1963} $I_{\rm eq}(\varphi)
= I_{\rm C} \sin\varphi$, with $I_{\rm C} = (e\Delta \tau/2\hbar) \tanh (\Delta/
2k_{\rm B}T )$. At perfect transparency ($\tau=1$), this expression reproduces
the Kulik-Omelyanchuk formula:\cite{Kulik1977} $I_{\rm eq}(\varphi) = I_0
\sin (\varphi /2) \tanh (\Delta \cos (\varphi /2)/ 2k_{\rm B}T )$. Here,
$I_0 = e\Delta/ \hbar$ is the zero-temperature critical current for $\tau=1$
and we  frequently use it below to normalize the supercurrent
in the different graphs. According to Eq.~(\ref{Ieq}), at zero temperature only
the lower ABS contributes to the supercurrent $I_{\rm eq}=I^+_{\rm A}$, while at
a finite temperature the negative contribution from the upper ABS leads to a
decrease of the total supercurrent.

\begin{figure}[t]
\begin{center}
\includegraphics[width=\columnwidth]{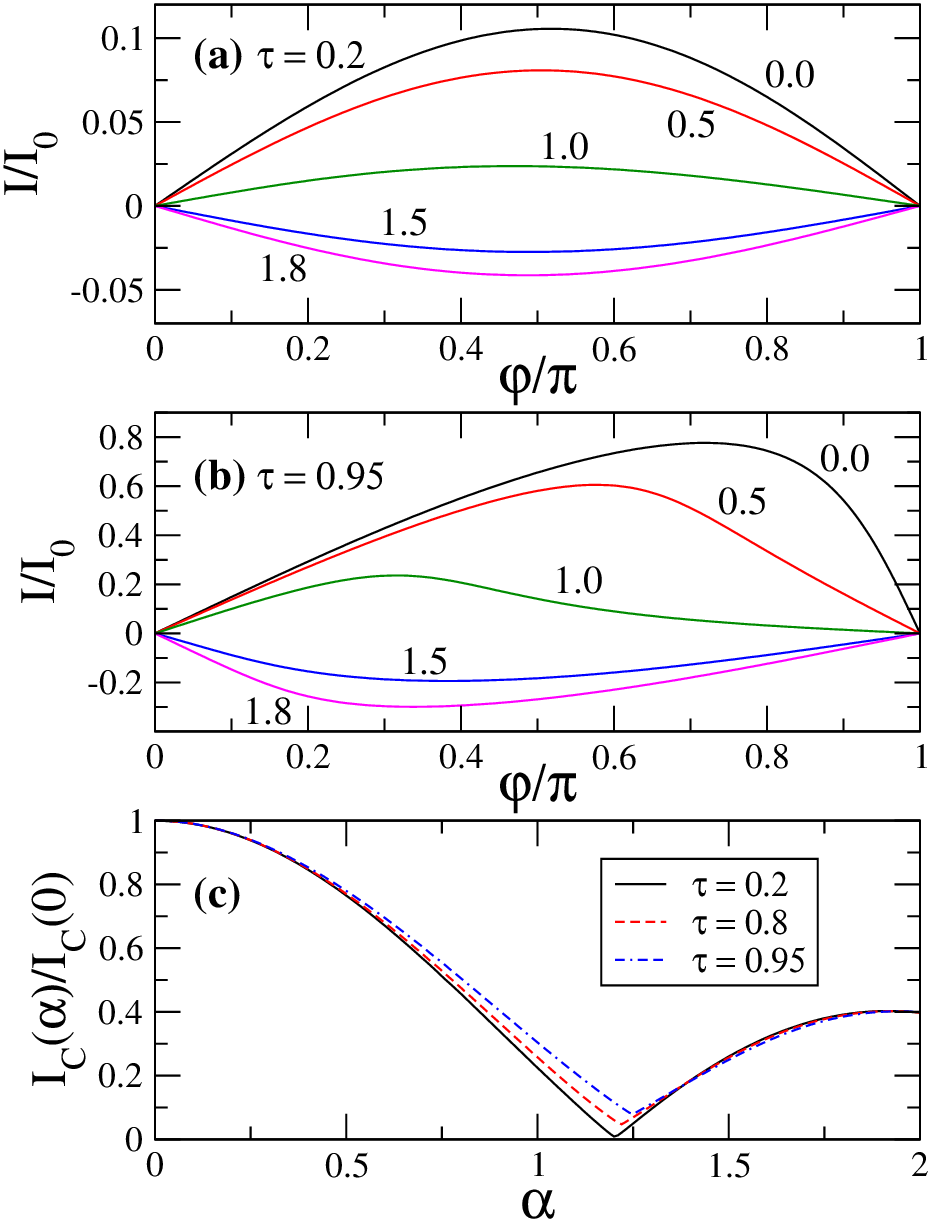}
\caption{(Color online) Panels (a) and (b): The current-phase relation in the
adiabatic approximation for $\tau=0.2$ (a) and $\tau=0.95$ (b). The different
curves correspond to different values of $\alpha$ as indicated in the graphs.
The current is given in units of $I_0=e\Delta/\hbar$, where $\Delta$ is the value
of the superconducting gap at $T=0$. (c) The zero-temperature critical current
as a function of $\alpha$ for three different values of the transmission $\tau$.
Notice that critical current is normalized by its value in the absence of
microwaves.\label{fig_Iad}}
\end{center}
\end{figure}

The simplest approach to compute the supercurrent in the presence of the
microwave field is the  so-called adiabatic approximation.\cite{Barone1982}
In this approximation one assumes that the ABSs follow adiabatically the ac
drive and there are no direct transitions between them. Thus, the CPR in
this approximation is obtained by replacing the stationary phase $\varphi$
in Eq.~(\ref{Ieq}) by the time-dependent phase $\phi(t)$ of Eq.~(\ref{eq-phi}),
which leads to the following result
\begin{equation}
I_{\rm ad}(\varphi,\alpha) = \sum_{n=1}^{\infty}I_n J_0(2n\alpha)
\sin(n\varphi), \label{Iad}
\end{equation}
where $I_n=(1/\pi) \int_0^{2\pi} d\varphi \; I_{\rm eq}(\varphi) \sin(n\varphi)$
are the harmonics of the equilibrium CPR of Eq.~(\ref{Ieq}) and $J_0$ is
the zero-order Bessel function of the first kind. Notice that the current in this
approximation does not depend explicitly on the radiation frequency. We illustrate
the results of this approximation in Fig.~\ref{fig_Iad} for the zero-temperature case.
In particular, in the two upper panels we show the CPR (obtained from Eq.~\eqref{Iad})
for two different transmissions and several values of the $\alpha$ parameter
(related to the microwave power). Panel (a) corresponds to the tunnel limit
($\tau=0.2$) where the CPR is sinusoidal irrespective of the radiation power,
while in panel (b) we show the results for a high transmission of $\tau=0.95$.
In this latter case, the critical current is reached at different values of the
phase depending on the value of $\alpha$. Notice that no matter the value of the
phase $\varphi$, the magnitude of the supercurrent is always suppressed by the
microwaves as compared with the zero-field result ($\alpha=0$), which is true at
any temperature. With respect to the behavior of the critical current $I_C(\alpha)$,
as one can see in Fig.~\ref{fig_Iad}(c), it decays in a non-monotonic manner, as
governed by the Bessel function $J_0$.

\section{The two-level model}\label{TLM}

It is instructive to start our analysis towards a microscopic theory by
restricting ourselves to the study of the contribution of ABSs, ignoring
for the moment the continuum part of the spectrum. This can be done with
the help of the two-level models that have been derived in
Refs.~\onlinecite{Ivanov1999} and \onlinecite{Zazunov2003} to describe the
dynamics of a SQPC under external ac fields. The models of these two references
coincide at equilibrium, but they differ slightly when the phase depends
on time. In particular, the model of Ref.~\onlinecite{Zazunov2003} ensures
charge neutrality, while the model of Ref.~\onlinecite{Ivanov1999} does
not. For this reason, we  base our discussion here on the model put
forward by Zazunov and coworkers.\cite{Zazunov2003} In this model, the
SQPC is described by the $2 \times 2$ Hamiltonian
\begin{equation}
\hat H_{\rm B}(t) = \Delta e^{-i\hat\sigma_x r\phi/2}
\left(\cos\frac{\phi}{2} \hat\sigma_z +
r \sin\frac{\phi}{2}\hat\sigma_y\right)\;,\label{TLH-B}
\end{equation}
where $r=\sqrt{1-\tau}$ and $\phi(t)$ is the time-dependent phase given by
Eq.~(\ref{eq-phi}). This Hamiltonian is written in the ballistic basis of
right- and left moving electrons, which are eigenvectors of the current operator
in the perfectly transmitting case ($\tau=1$). For our subsequent analysis it is
more convenient to work in the instantaneous Andreev basis $\{\ket{+}_{\phi(t)}$,
$\ket{-}_{\phi(t)}\}$, whose basis vectors are time-dependent. This is the basis
where the Hamiltonian of Eq.~(\ref{TLH-B}) becomes diagonal in equilibrium. The
Andreev basis is obtained from the ballistic basis by means of a transformation
$\hat H_A(t)=\hat R^\dagger (t)\hat H_B(t)\hat R(t)$ generated by the  unitary matrix
\begin{equation}
\hat R(t) = e^{-i \hat \sigma_x r\frac{\phi}{4}} e^{-i\frac{\pi}{4}\hat\sigma_z}
e^{-i\theta(\phi)\hat\sigma_y} ,
\end{equation}
where $\theta(\phi)=(1/2) \arctan [r \tan(\phi/2) ]$. With this transformation
the Schr\"{o}dinger equation for a state vector $\Psi(t)=(\alpha(t),\beta(t))^T$
becomes
\begin{equation}
i \partial_t \Psi(t)=\hat H_{\rm A}(t) \Psi(t)\; \label{schr},
\end{equation}
where
\begin{equation}
\hat H_{\rm A}(t) = E_{\rm A}(\phi(t))\hat{\sigma}_z -
\frac{r \tau \Delta^2 \sin^2(\phi(t)/2)} {4[E_{\rm A}(\phi(t))]^2}
\dot{\phi}(t) \hat{\sigma}_y , \label{tlm}
\end{equation}
and $\dot{\phi}(t) = \partial \phi(t) / \partial t$. Moreover, in
the previous two equations, and in the rest of this section, we set $\hbar = 1$.

The corresponding current operator can be written as
\begin{equation}
\hat I_{\rm A}(t) = 2eE^{\prime}_{\rm A}(\phi(t))\hat{\sigma}_z +
\frac{e r \tau \Delta^2 \sin^2(\phi(t)/2)}
{E_{\rm A}(\phi(t))} \hat{\sigma}_x ,  \label{ctlm}
\end{equation}
where the prime in $E^{\prime}_{\rm A}$ means derivative with respect to the
argument (the time-dependent phase in this case).
To obtain the expectation value of the current at different
times, Eq.~\eqref{schr} needs to be solved. Despite the apparent simplicity,
this task has nontrivial aspects: straightforward numerical approaches
run into problems, as both very fast ($t^{-1}\sim{}\omega$) and very slow
($t^{-1}\sim{}E_{\rm A} - n\omega$) time scales can be simultaneously present.
No closed-form analytical solution can be obtained either,\cite{Grifoni1998}
and the significantly nonlinear coupling to the drive makes it more difficult
to derive approximations via standard routes.\cite{Bloch1940,Autler1955}

Focusing our analysis on time-averaged quantities, we can obtain accurate
analytical and numerical results via a systematic Floquet-type approach.
We are interested in two physical quantities: the dc current
\begin{equation}
\bar I=\lim_{t\to\infty}\frac{1}{t}\int_0^tdt'\Psi^\dagger(t') \hat I(t')\Psi(t')
\,, \label{avr_current}
\end{equation}
and the time-averaged populations of the Andreev levels
\begin{equation}
\bar p_\pm = \lim_{t\to\infty} \frac{1}{t} \int_0^t dt^{\prime}
\Psi^\dagger(t^{\prime}) \frac{\hat 1 \pm \hat \sigma_z}{2} \Psi(t^{\prime})
\label{avr_pop}\; .
\end{equation}
Below, we show how to obtain $\bar I$, although the method described
can as well be used to compute any other time-averaged quantity,
including $\bar p_\pm$.

We first introduce a modified Hamiltonian
\begin{equation}
  \hat H_{\rm A}(t,\chi)=\hat H_{\rm A}(t)+\chi\hat I_{\rm A}(t)\; ,\label{H_chi}
\end{equation}
where $\chi$ is a parameter conjugate to the observable, and it is set to
zero at the end of the calculation. The solution of the Schr\"{o}dinger
equation $\Psi(t,\chi)$ can be formally written by introducing the time evolution
operator $\hat U(t,0;\chi)$
\begin{equation}
\label{sol_schr}
\Psi(t,\chi)={\cal T} e^{-i\int_0^t dt'\hat H_A(t',\chi)}\Psi_0\equiv
\hat U(t,0;\chi)\Psi_0 \,,
\end{equation}
where ${\cal T}$ indicates time ordering and $\Psi_0$ is the state
vector at $t=0$. We define now the generating function:
\begin{equation}
  S(t,\chi)=\Psi_0^\dagger U(0,t;\chi=0)U(t,0;\chi)\Psi_0\,. \label{gen_func}
\end{equation}
One can easily check that the dc current defined in Eq.~\eqref{avr_current}
can be written as
\begin{equation}
\bar I=\lim_{t\to\infty} \frac{i}{t}\partial_\chi S(t,\chi)|_{\chi=0}\;.
\label{dc_curr}
\end{equation}
Thus, we only need to compute the function $S$, or, equivalently, the evolution
operator $\hat U(t,0;\chi)\equiv \hat U(t;\chi)$.

Since our Hamiltonian is periodic in time with a period $T=2\pi/\omega$, i.e.,
$H_{\rm A}(t,\chi)=H_{\rm A}(t+T;\chi)$, we can define two periodic (Floquet)
states ${\rm v}_{\pm}$ via the eigenvalue problem
\begin{equation}
  \hat U(T;\chi){\rm v}_{\pm}(\chi)=e^{\pm iE(\chi)T}{\rm v}_{\pm}(\chi)
  \,.
\end{equation}
The symmetry of the two eigenvalues follows here from the fact that $\hat
H_{\rm A}(t,\chi)$ and ${\rm log}[U(T;\chi)]$ are traceless, and
$U(T;\chi)$ is unitary.  Moreover, from the periodicity of the
Hamiltonian it follows that
\begin{equation}
\hat U(nT;\chi)=\hat U(T;\chi)^n=\hat V(\chi)e^{iE(\chi)nT\hat\sigma_z}\hat V^{-1}(\chi)\;,
\end{equation}
where the eigenvectors ${\rm v}_{\pm}$ form the columns of the unitary
matrix $\hat V$. Replacing in Eqs.~(\ref{gen_func}) and (\ref{dc_curr}) $t$
by $nT$ and taking the limit $n\rightarrow\infty$ we now find the
derivative with respect to $\chi$:
\begin{equation}
\frac{1}{nT}\partial_\chi \hat U(nT;\chi)\stackrel{n\rightarrow\infty}
{\longrightarrow}i\hat V^{-1}(\chi)\hat\sigma_ze^{iE(\chi)nT\hat\sigma_z}
\hat V(\chi)\frac{\partial E(\chi)}{\partial \chi} \,.
\end{equation}
Thus, the dc current is given by
\begin{equation}
\bar I = -\Psi_0^\dagger \left( {\rm v}_+{\rm v}_+^\dagger-{\rm v}_-{\rm v}_-^\dagger
\right) \Psi_0 \left. \frac{\partial E(\chi)}{\partial \chi}\right|_{\chi=0}
\; .\label{Idc_num}
\end{equation}
This exact expression for the dc current is very useful for numerics. It is
easy to compute and it handles the fast and slow time scales of the problem
separately. In order to obtain the dc current, one needs first to integrate
the Schr\"{o}dinger equation with the Hamiltonian of Eq.~(\ref{H_chi}) over one
period to find the $2\times2$ matrix $\hat U(T;\chi)$, then one computes its
eigenvalues $\pm{}E$ and eigenvectors ${\rm v}_\pm$, and finally the derivative
$\partial_\chi{}E(\chi)$ is computed via numerical differentiation.

\begin{figure*}[t]
\begin{center}
\includegraphics[width=0.9\textwidth,clip]{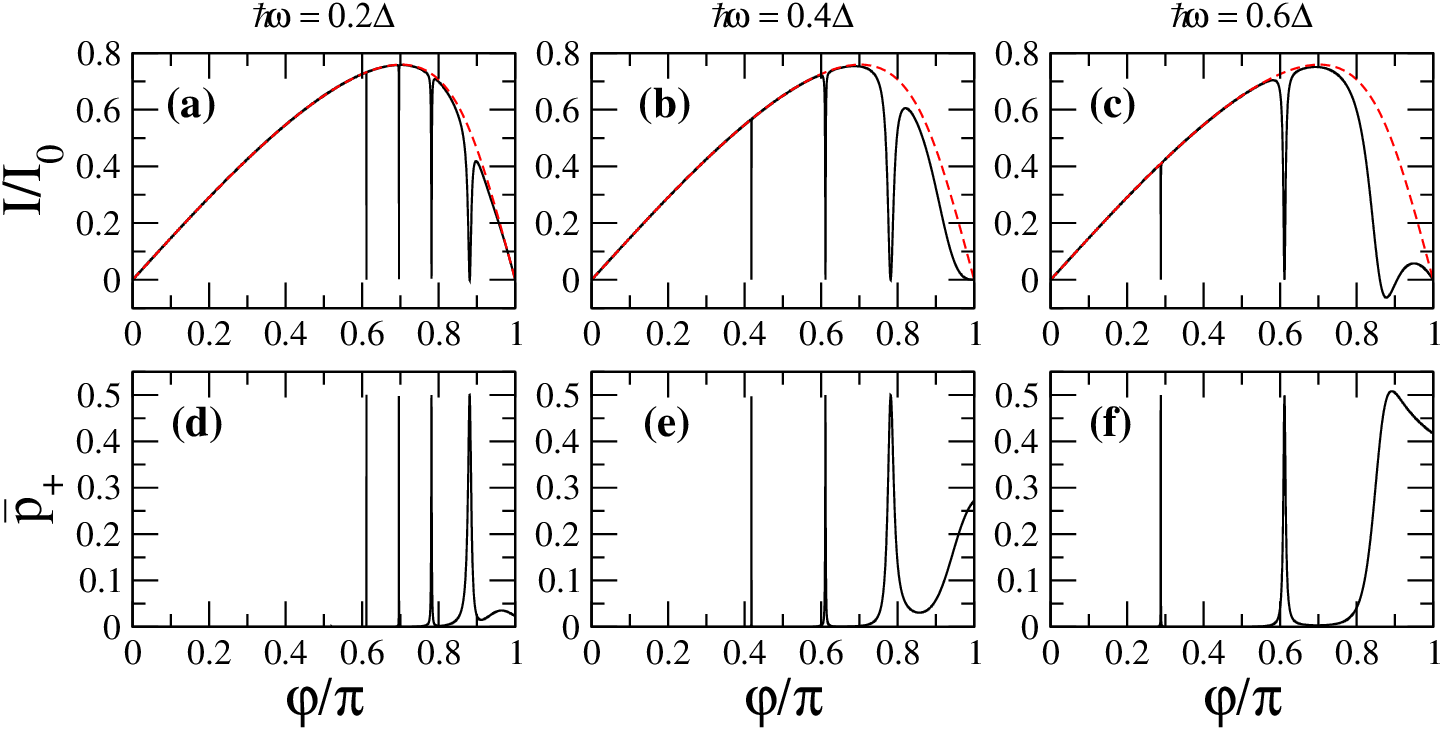}
\caption{(Color online) (a-c) Zero temperature supercurrent, in units of $I_0 =
e\Delta/\hbar$, as a function of the phase for $\tau=0.95$, $\alpha=0.15$ and
three different values of the microwave frequency, as indicated in the upper
part of the graphs. The solid lines correspond to the numerical results obtained
with the two-level model, while the dashed lines is the result obtained with
the adiabatic approximation. (d-f) Time-average occupation of the upper ABS for
the cases shown in the upper panels.} \label{2level-numerics}
\end{center}
\end{figure*}

In order to have a first impression of the results from this two-level model,
we show in Fig.~\ref{2level-numerics} a few examples of the CPR of a highly
transmissive channel ($\tau=0.95$) computed with the numerical recipe that
we have just described.\cite{note-IC} The upper panels of this figure show the CPR
for a moderate power ($\alpha = 0.15$) and three different values of the microwave
frequency. For comparison, we also show the result obtained with the adiabatic
approximation of Eq.~(\ref{Iad}). As one can see, the main difference is the
appearance in the results of the two-level model of a series of dips at certain
values of the phases where the current is largely suppressed. It is easy to
understand that such dips are due to microwave-induced transitions between the ABSs.
These transitions enhance the population
of the upper ABS, which at zero temperature would be empty otherwise, and at the
same time they reduce the occupation of the lower ABS. This redistribution of
the quasiparticles in turn results in a suppression of the current. The
microwave-induced transitions occur with the highest probability when the
distance in energy between the ABSs (the Andreev gap) is equal to a multiple
of the photon energy, i.e., when $2E_{\rm A}(\varphi) = n\omega$, where
$n=1,2,\dots$ is the number of photons involved in the transition. If this
condition is expressed in terms of the phase $\varphi$, it adopts the 
from
\begin{equation}
\varphi_n = 2\, {\rm arcsin} \sqrt{[1-(n \omega/2\Delta)^2]/\tau},
\; \; n= 1,2,\dots\; . \label{phin}
\end{equation}
A detailed analysis shows that this expression reproduces the positions of all
the dips appearing in the examples of Fig.~\ref{2level-numerics}.

This interpretation of the origin of the dips in the CPR can be corroborated
by a direct analysis of the occupations of the ABSs. Following the same
numerical recipe, we have also computed the average occupation of the upper ABS,
$\bar{p}_+$, for the examples shown in the upper panels of Fig.~\ref{2level-numerics}.
The results can be seen in the lower panels of this figure and, as one can observe,
there is a clear one-to-one correspondence between the current dips and the
enhancement of the time-averaged population of the upper state. In particular, whenever the
upper state reaches a population equal to $1/2$, the current vanishes exactly.

The method described above is not only very convenient for numerical calculations,
but it also provides a route to obtain analytical results. In what follows, we 
show how this method can be used, in particular, to gain a further insight into
the microwave-induced supercurrent dips. To proceed,
it is useful to first rewrite Eq.~\eqref{Idc_num} in a more convenient form. In particular,
we would like to avoid the calculation of eigenvectors in this equation. This can
be done by noting that the unperturbed Hamiltonian~\eqref{tlm} obeys
\begin{align}
\hat \sigma_x\hat H_{\rm A}\hat\sigma_x=-\hat H_{\rm A}\; .
\end{align}
Consequently, ${\rm v_-}\propto{}\sigma_x{\rm v_+}$, and the dc current given
by Eq.~\eqref{Idc_num} can be written as
\begin{equation}
\bar I = -{\rm v}_+^\dagger \left(\hat\rho_0-\hat\sigma_x\hat\rho_0\hat\sigma_x \right)
{\rm v}_+\left.\frac{\partial E(\chi)}{\partial \chi}\right|_{\chi=0}\; ,\label{Idc_num2}
\end{equation}
where $\hat \rho_0 = \Psi_0\Psi_0^\dagger$. Using the expression for the
change of an eigenvalue due to a perturbation, we can finally write
\begin{equation}
\bar I=\left. \frac{\partial E(\chi,\mu)}{\partial \mu}\frac{\partial E(\chi,\mu)}
{\partial \chi}\right|_{\chi,\mu=0}\; .\label{Idc_ana}
\end{equation}
Here $ E(\chi,\mu)$ is an eigenvalue of the matrix
\begin{equation}
\hat M(\chi,\mu) = \frac{i}{T}\hat\Omega(T,\chi) +
\mu\left(\hat\rho_0-\hat\sigma_x\hat\rho_0\hat\sigma_x\right)\;, \label{matrixM}
\end{equation}
$\hat\Omega(T,\chi)\equiv{\rm log}[U(T,\chi)]$ and $\mu$ is an additional
perturbation parameter. The problem is now reduced to finding the eigenvalues
of a 2$\times$2 matrix.

As discussed above, the dc current for weak fields only deviates from the
adiabatic result close to the resonant conditions $n\omega = 2E_{\rm A}$ (with
$n=1,2,3,\ldots$), where the transitions between the ABSs are more likely.
In order to study what happens close to these resonant situations, we can
consider the problem in a rotating frame, and rewrite the evolution operator
$\hat U(t)$ defined in Eq.~\eqref{sol_schr} as
\begin{equation}
\hat U(t)=e^{-i\hat W_nt}{\cal T}e^{-i\int_0^tdt^{\prime} \hat{\tilde H}_n(t^{\prime})}
\equiv e^{-i\hat W_nt}\hat{\tilde U}_n(t) \,, \label{int_pic}
\end{equation}
where $\hat W_n=n\omega \hat \sigma_z/2$ and the rotating-frame Hamiltonian
is
\begin{equation}
\hat{\tilde H}_n(t)=e^{i\hat W_nt}[\hat H_{\rm A}(t)-\hat W_n]
e^{-i\hat W_nt}\label{new_H}\;.
\end{equation}
The generating function can then be written as in Eq.~\eqref{gen_func}
simply by substituting $\hat{H}_{\rm A}$ by $\hat{\tilde H}_n$. The
additional exponential factors simply cancel out, and the Hamiltonian
$\hat{\tilde H}_n(t)$ remains periodic. Thus, we can proceed exactly as
above.

The key idea that allows us to obtain analytical results is the fact that
for weak fields ($\alpha\ll1$), the dynamics in the rotating frame are slow
($\hat{\tilde{H}}_n$ is small) around the corresponding resonance. For this
reason, we can use the Magnus expansion\cite{Magnus54} to determine the matrix
$\hat \Omega$ appearing in Eq.~\eqref{matrixM}:
\begin{align}
  \label{magnus}
  \hat \Omega(T)=&-i\int_0^T dt_1\hat H_n(t_1)
  \\\notag
  &-\frac{1}{2} \int_0^T dt_1\int_0^{t_1}dt_2\left[\hat H_n(t_1),\hat H_n(t_2)\right]+\ldots
  \,.
\end{align}
This is essentially an expansion in the parameter $\lambda_n \sim 2n\pi
(E_{\rm A}-n\omega/2)/\omega$, which indeed is small close to a resonance.

We proceed now computing the dc current close to the first resonance
$\omega=2E_{\rm A}$, assuming that initially the system is in its ground
state $\Psi_0^\dagger=(0,1)$. We choose $\hat W=\omega\sigma_z/2$, and take
only the first term of the expansion~\eqref{magnus}, expanding up to the first
order in $\alpha$ and $\chi$. The time integral is straightforward to evaluate,
and we obtain
\begin{align}
  \frac{i}{T}
  \hat{\tilde \Omega}_1(T,\chi)
  &\simeq
    \bigl[ E_{\rm A}-\frac{\omega}{2}+2e\chi E_{\rm A}'\bigr]\hat\sigma_z
  \\\notag
  &
  +
  \frac{r}{2E_{\rm A}^2}
  \alpha\bigl[
    (\Delta^2-E_{\rm A}^2)\frac{\omega}{2}-\chi (\Delta^2+E_{\rm A}^2) 2eE_{\rm A}'\bigr]
    \hat\sigma_x\,.
\end{align}
Note that this expression is analogous with the well-known rotating
wave approximation, with the difference that by considering the
generating function, our formalism takes the time dependence of the
operator $\hat{I}(t)$ into account.  For the eigenvalues of the matrix
$\hat{\tilde M}_1=(i/T) \hat{\tilde \Omega}_1+\mu\sigma_z$ we obtain
\begin{align}
  E^2 &= [E_{\rm A} - \frac{\omega}{2} + \mu + 2e\chi E_{\rm A}']^2
  \\\notag
  &+(r\alpha/2E_{\rm A}^2)^2[
  (\Delta^2-E_{\rm A}^2)\frac{\omega}{2}-
  \chi (\Delta^2+E_{\rm A}^2) 2e E_{\rm A}']^2
  \,.
\end{align}
Finally, working in the limit $(\omega-2E_{\rm A})/\Delta\ll1$ for simplicity,
we find the dc current from Eq.~\eqref{Idc_ana}:
\begin{equation}
 \bar I _1(\varphi,\omega,\alpha)
  \approx
   -2eE_{\rm A}'\left(
    1
    -
    \frac{
      \Omega_{r,1}^2
    }{
      (\omega-\omega_1)^2 + \Omega_{r,1}^2
    }
  \right)
  \,,\label{I1}
\end{equation}
where the resonant frequency $\omega_1 = 2E_{\rm A}$ equals the unperturbed
Andreev level spacing $2E_{\rm A}$ (up to first order in $\alpha$), and
$\Omega_{r,1} = r \alpha \omega (\Delta^2-E_{\rm A}^2)/2E_{\rm A}^2$ is the
corresponding Rabi frequency. This expression tells us that the current
vanishes exactly at the resonant condition $\omega = \omega_1$ and that
the width of the current dip is given by $\Omega_{r,1}$, which is linear
in $\alpha$. Moreover, its form clearly suggests that the populations of
the two states undergo Rabi oscillations with the frequency $\Omega_{r,1}$, as
usual in two-state systems, and the time-averaged populations of the
ABS coincide at the resonance. As a consequence, the dc current drops
to zero at the resonance, a result that qualitatively coincides with the
prediction in Ref.~\onlinecite{Shumeiko1993}.

We can also determine the dc current at the higher resonances, for
example for $\omega\approx E_A$. In this case we work in the frame
corresponding to $\hat W_2=\omega\hat\sigma_z$. As the resonance is
due to two-photon processes, terms up to order $\alpha^2$ must be
taken into account, which requires including the first two terms in
Eq.~\eqref{magnus}. The computations are again straightforward, and up
to the second order in $\alpha$ we obtain
\begin{align}
  \frac{i}{T}
  \hat{\tilde \Omega}_2
  &
  \approx \bigl[E_{\rm A} - \omega + \alpha^2E_{\rm A}'' + 2e\chi E_{\rm A}'
  +
  r^2 \alpha^2\omega\frac{(\Delta^2-\epsilon^2)^2}{12E_{\rm A}^4}\bigr]\hat\sigma_z
  \\\notag
  &
  - \frac{r\alpha^2E_{\rm A}'}{2E_{\rm A}^2}
\bigl[ \Delta^2\bigl(\frac{\omega}{E_{\rm A}}+1\bigr) -E_{\rm A}^2)\bigr]\hat\sigma_x\; .
\end{align}
For simplicity, we dropped terms of order $\alpha \chi$, which do not
essentially affect the form of the resonance. As above, the current is
obtained from the eigenvalues of $\hat{\tilde M}_2=(i/T) \hat{\tilde
\Omega}_2+\mu\sigma_z$, and it adopts the form
\begin{equation}
\bar I_2(\varphi,\omega,\alpha) \simeq-2eE_{\rm A}'\left(1-\frac{\Omega_{r,2}^2}
{(\omega-\omega_{2})^2+\Omega_{r,2}^2}\right) \,,\label{I2}
\end{equation}
where $\omega_{2} = E_{\rm A} + \alpha^2E_{\rm A}'' +
r^2\alpha^2(\Delta^2-E_{\rm A}^2)^2/12E_{\rm A}^3$ and $\Omega_{r,2} =
r \alpha^2(2\Delta^2- E_{\rm A}^2)E_{\rm A}'/2E_{\rm A}^2$.  Here, one can
observe that the resonant frequency is shifted from the position
$2\omega=2E_{\rm A}$ by two contributions: the first arises from nonlinearities,
and the second is the Bloch-Siegert shift.\cite{Bloch1940}

One can also go further and compute the dc current around resonances
$n>2$, although this gets progressively more cumbersome as an
increasing number of terms are required in the Magnus expansion,
reflecting the increasing number of allowed multiphoton processes
generated by the nonlinearities. One can however see from
Eqs.(\ref{I1},\ref{I2}), and also check for higher resonances, that
the width of the resonances scales with $\alpha^n$. Moreover,
one can show that within this model, the time-averaged current
vanishes exactly at each resonance, i.e., $\bar{I}\rvert_{\omega=\omega_n}=0$.

\begin{figure}[t]
\begin{center}
\includegraphics[width=0.9\columnwidth,clip]{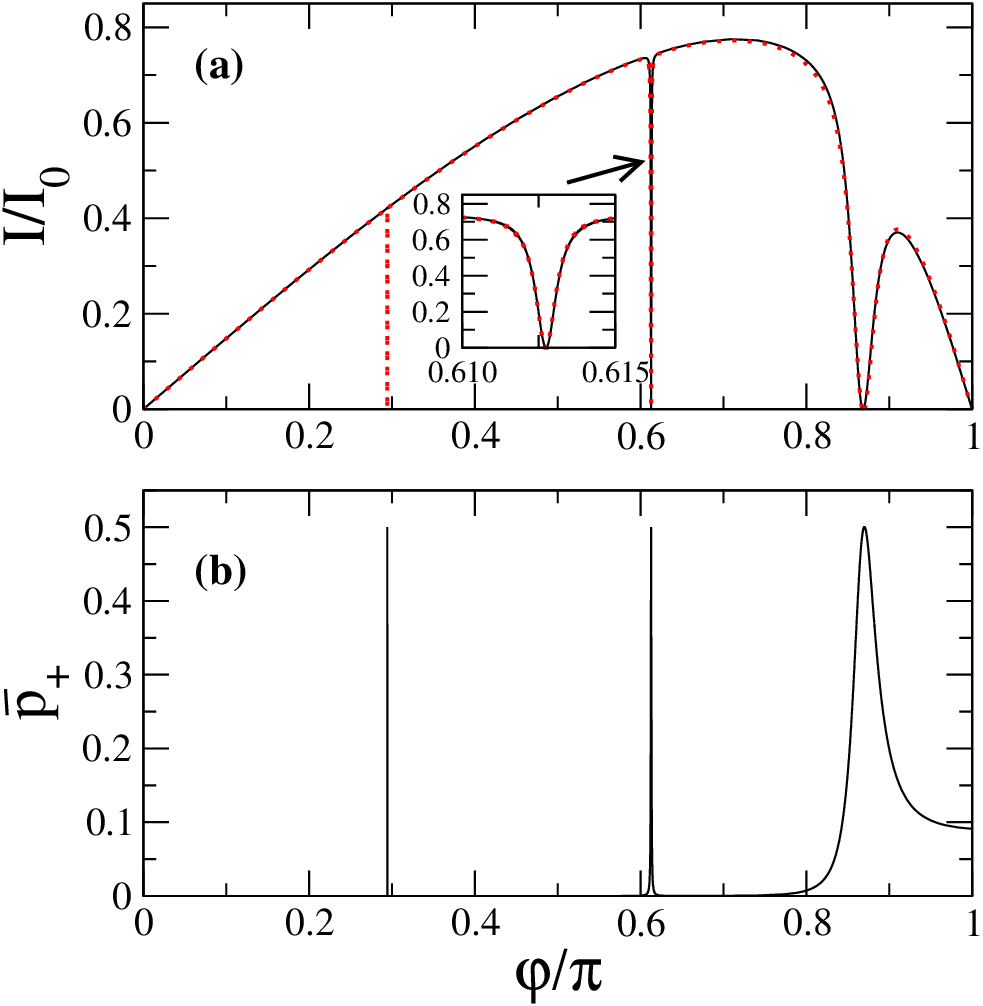}
\caption{\label{tlm-cpr}
(Color online) (a) Current-phase relation in the two-level model,
obtained from Eq.~\eqref{eq:tlm-compound-approx} (solid line) and
numerics based on Eq.~\eqref{Idc_num} (dotted line). Parameters are
$\omega=0.6\Delta$, $\tau=0.95$, and $\alpha=0.05$. The $n=3$ resonance
is not included in the analytical approximation. Inset: close-up of the
second resonance. (b) Time-averaged population $\bar{p}_+$ of the upper
Andreev state for the same parameters.}
\end{center}
\end{figure}

The results Eqs.(\ref{I1},\ref{I2}) can be combined into a single
approximate expression
\begin{equation}
  \label{eq:tlm-compound-approx}
  \bar I
  \approx
  -2eE_A'
  \biggl(
    1 - \frac{\Omega_{r,1}^2}{(\omega-\omega_{1})^2 + \Omega_{r,1}^2}
  \biggr)
  \biggl(
    1 - \frac{\Omega_{r,2}^2}{(\omega-\omega_{2})^2 + \Omega_{r,2}^2}
  \biggr)
  \,.
\end{equation}
The quality of this approximation can be established by comparing it with
the exact numerical results. This is done in Fig.~\ref{tlm-cpr} where
we have considered the case of a weak microwave field ($\alpha=0.05$).
 There is an excellent agreement between Eq. (\ref{eq:tlm-compound-approx}) and  the numerical results, apart from the fact that the numerics also include a dip
produced by three-photon processes, which we have left out from the
above approximation.

We can conclude this section by saying that in spite of the simplicity
of the two-level model considered here, such a model captures the essential
physics of the microwave-irradiated SQPC and, as we show in the next section, 
 it provides accurate results for not too high frequencies and up to moderate radiation power. 
 As we establish in the next section, the limitations of the two-level
model are mainly related to the fact that it does not take into account the
contribution of  the continuum of states outside the gap region.

\section{The Keldysh-Green function approach}\label{MM}

In the previous section we have analyzed the supercurrent assuming that the only
contribution comes from the ABSs. While this is true for a SQPC in equilibrium,
it is not obvious that this should be the case in the presence of a microwave
field. Indeed, at high frequencies or at high radiation powers, and especially
at finite temperatures, transitions between the ABSs and the continuum of states
outside the gap become possible and, in principle, they can also contribute to the
current. Therefore, in order to describe the complete phenomenology of irradiated
SQPCs we must develop a fully microscopic theory. This is the goal of this section.

Our microscopic theory is based on the Keldysh-Green function approach.
In this approach the starting point is the expression for the quasiclassical
Green functions of the left (L) and right (R) electrodes. In our case, these
Green functions can be expressed in terms of the equilibrium Green functions
$\check g(t-t^{\prime})$ as 
\begin{equation}
\check G_{R (L)}(t,t^{\prime}) = e^{\pm i\phi(t) \hat \tau_3/2}
\check g(t-t^{\prime}) e^{\mp i\phi(t^{\prime})\tau_3/2}\; .
\label{eq-GF-leads}
\end{equation}
Here $\phi(t)$ is the time-dependent phase given by Eq.~(\ref{eq-phi}) and
the upper (lower) sign in the exponents corresponds to the R (L) electrode.
The symbol $\check{\,}$ indicates that the Green functions are 4$\times$4
matrices in the Keldysh-Nambu space, where they have the structure
\begin{equation}
\check G=\begin{pmatrix}
\hat G^R & \hat G^K\\0 & \hat G^A
\end{pmatrix}\; .
\end{equation}
Here the symbol $\hat{\,}$ indicates that the different elements are 2$\times$2
Nambu matrices. The retarded (R), advanced (A) and Keldysh (K) components
of the equilibrium Green functions appearing in Eq.~(\ref{eq-GF-leads})
are given by
\begin{equation}
\check g(t)= \int^{\infty}_{-\infty} \frac{dE}{2\pi} e^{-iE t/\hbar} \check g(E) ,
\end{equation}
where
\begin{eqnarray}
\hat g^{R(A)}(E)& =& g^{R(A)}(E) \hat \tau_3 + f^{R(A)}(E) i \hat \tau_2\\
\hat g^{K}(E) &=& \left[\hat g^{R}(E) - \hat g^{A}(E)\right] \tanh(E/2k_{\rm B}T)
\end{eqnarray}
and
\begin{equation}
g^{R(A)}(E) = \frac{E}{\sqrt{(E \pm i\eta)^2-\Delta^2}} =
\frac{E}{\Delta} f^{R(A)}(E) ,
\end{equation}
where $\eta$ describes the inelastic scattering energy rate within the relaxation
time approximation and $T$ is the temperature.

Different authors have shown that the transport properties of a point contact
with an arbitrary time-dependent voltage can be described by making use of
adequate boundary conditions for the full quasiclassical
propagators.\cite{Zaitsev1998,Nazarov1999,Cuevas2001,Kopu2004} These boundary
conditions can be expressed in terms of a current matrix
\begin{equation}
\check I= \left( \begin{array}{ccc}
\hat I^R & \hat I^K  \\
0 & \hat I^A \end{array} \right) ,
\end{equation}
which for the case of a single-channel SQPC of transmission $\tau$ can be
expressed in terms of the lead Green functions of Eq.~(\ref{eq-GF-leads})
as\cite{Nazarov1999}
\begin{equation}
\check I(t,t^{\prime}) = 2 \tau \left[\check G_L , \check G_R \right]_{\circ}
\circ \left[ 4 - \tau \left(2- \left\{ \check G_L , \check G_R \right\}_{\circ}
\right) \right]^{-1} (t,t^{\prime})  \label{matrixJ}.
\end{equation}
Here, the symbol $\circ$ denotes the convolution over intermediate time arguments.
Finally, the electric current is obtained by taking the trace
\begin{equation}
I(t) = \frac{e}{4\hbar} {\rm Tr}\hat \tau_3 \hat I^K(t,t)\; ,
\end{equation}
where $\hat \tau_3$ is the third Pauli matrix in Nambu space.

Due to the periodic time dependence of the phase [Eq.~(\ref{eq-phi})], the
Green functions $\check G_{L(R)}$, and any products of them, admit the following
Fourier expansion
\begin{equation}
\check G(t,t^{\prime}) = \sum^{\infty}_{m=-\infty} e^{im\omega t^{\prime}}
\int \frac{dE}{2\pi} e^{-i E(t-t^{\prime})/\hbar} \check G_{0m}(E) ,
\end{equation}
where $\check G_{nm}(E) \equiv \check G(E+n\hbar \omega, E+m\hbar \omega)$ are
the corresponding Fourier components in energy space, and $n,m$ are integers. In particular, the Fourier components of $\check G_{L(R)}$ can be
 deduced from Eq.~(\ref{eq-GF-leads}). For instance, for the left
electrode, $\check G_{nm}(E)$ is given by
\begin{equation}
(\check G_L)_{nm} = \sum_l \check \Gamma_{nl} \check g_l \check \Gamma^{*}_{lm} ,
\end{equation}
where
\begin{equation}
\check \Gamma_{nm} = \left(
\begin{array}{cc}
\hat \Gamma_{nm} & 0\\
 0 & \hat \Gamma_{nm} \end{array} \right) ,
\hat \Gamma_{nm} = \left( \begin{array}{cc}
{\cal P}_{nm} & 0\\
0 & {\cal P}^{\ast}_{nm} \end{array} \right) .
\end{equation}
Here, ${\cal P}_{nm}=(i)^{m-n}J_{m-n}(\alpha/2)e^{i\varphi/4}$, where $J_n$ is
the Bessel function of order $n$, and $\check g_n=\check g(E+n\hbar\omega)$ is
the equilibrium Green function matrix with the argument shifted in energy.

From this discussion, it is easy to understand that the current adopts the
 general expression
\begin{equation}
I(t) = \sum^{\infty}_{m=-\infty} I_m e^{im \omega t} ,
\end{equation}
which means that the current oscillates in time with the microwave frequency and
all its harmonics. These current components can be computed from the Fourier components
in energy space of $\check I$ in Eq.~(\ref{matrixJ}). From that equation, it is
straightforward to show that the Fourier components of $\hat I^K$ are given by
\begin{equation}
\hat I_{nm}^K = \sum_l[\hat A_{nl}^R \hat X_{lm}^K + \hat A_{nl}^K \hat X_{lm}^A] .
\label{eq-J}
\end{equation}
Here, we have defined the matrices $\check A_{nm} \equiv 2 \tau
[\check G_L,\check G_R]_{nm}$ and  $\check X_{nm} = [4\check 1-\tau(2-
\{\check G_L,\check G_R\})]^{-1}_{nm}$, which can be determined from the Fourier
components of $\check G_{L(R)}$. Once the components of $\hat I^K$ are obtained
from Eq.~(\ref{eq-J}), one can compute the current. We are only interested here
in the dc component, which reads
\begin{equation}
I(\varphi,\omega,\alpha) = \frac{e}{4\hbar} \int \frac{dE}{2\pi} {\rm Tr}
\hat \tau_3 \hat I^K_{00}(E,\varphi,\omega,\alpha) .
\label{Idc}
\end{equation}
The dc current can be calculated analytically in certain limiting cases: for example in
the absence of microwaves, where it reduces to Eq.~(\ref{Ieq}), in the tunnel
regime or for very weak fields. However, for arbitrary radiation power one needs
to evaluate Eq.~(\ref{Idc}) numerically. In the next subsections we present the
results for the dc current of this microscopic theory and we compare them with
those obtained from the two-level model of Section \ref{TLM}.

\subsection{Zero-temperature limit: Comparison with the two-level model}

We focus first on the analysis of the results of the exact theory at zero
temperature. This  allows us, in particular, to make a comparison with
the two-level model of Section III and to establish its range of validity.

\begin{figure}
\begin{center}
\includegraphics[width=\columnwidth,clip]{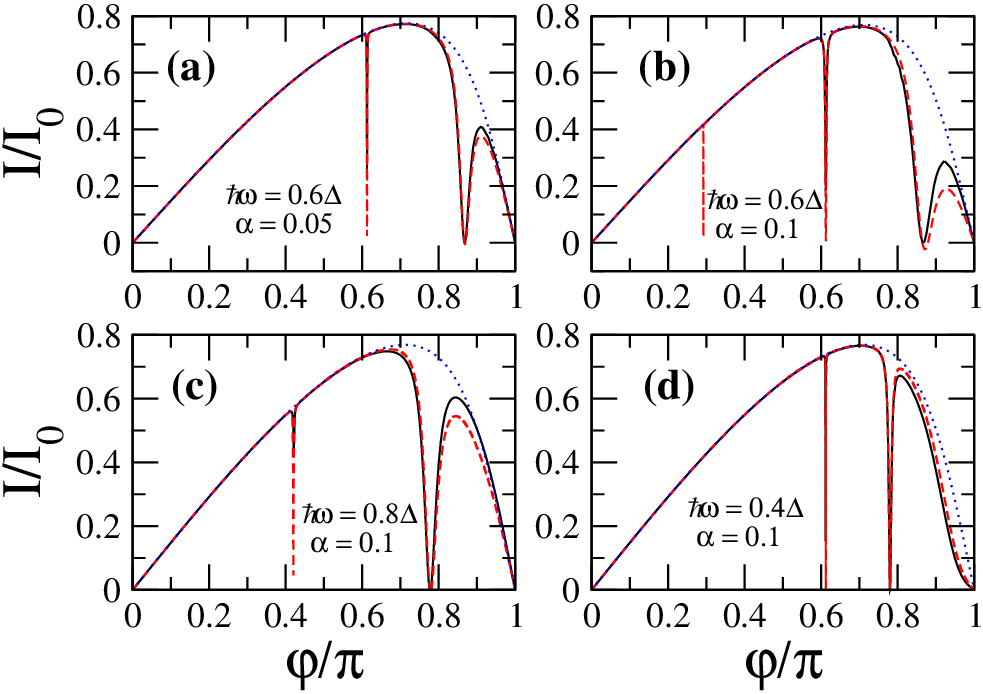}
\caption{\label{comparison1}
(Color online) Four examples of the zero-temperature current-phase relation for $\tau=0.95$
obtained from the microscopic model (solid lines), the two level model (dashed lines)
and the adiabatic approximation (dotted lines). The parameters characterizing the
microwave field are indicated in the different panels.}
\end{center}
\end{figure}

In Fig.~\ref{comparison1} we show several examples of the CPR calculated with
the microscopic approach (solid lines) for a highly conductive channel ($\tau=0.95$)
for several frequencies and low values of the radiation power ($\alpha \ll 1$). For
comparison, we also show the results of both the two-level model (dashed lines)
and the adiabatic approximation (dotted lines). As one can see, the main deviation
from the adiabatic results is the appearance of a series of dips, as discussed in
section III. These features, which originate from the microwave-induced transitions
between the ABSs, are accurately reproduced by the two-level model (both the position
and the width of the dips). There is a small discrepancy between the exact result
and those of the two-level model for phases close to $\pi$, i.e., when the level
spacing between the ABS is very small. This is understandable since the model
assumes that $\hbar \dot \phi(t) \ll 2E_{\rm A}$, which is not fulfilled when
$\varphi \sim \pi$ and $\tau$ is close to 1. Notice also that for the high-order
dips (due to high-order photonic processes), the suppression of the current in the two-level
model is larger than in the case of the exact theory. The reason is the additional
broadening introduced by the finite inelastic scattering rate used in the
calculations with the microscopic theory, which in this case is $\eta= 10^{-4}\Delta$.

\begin{figure}
\begin{center}
\includegraphics[width=0.8\columnwidth,clip]{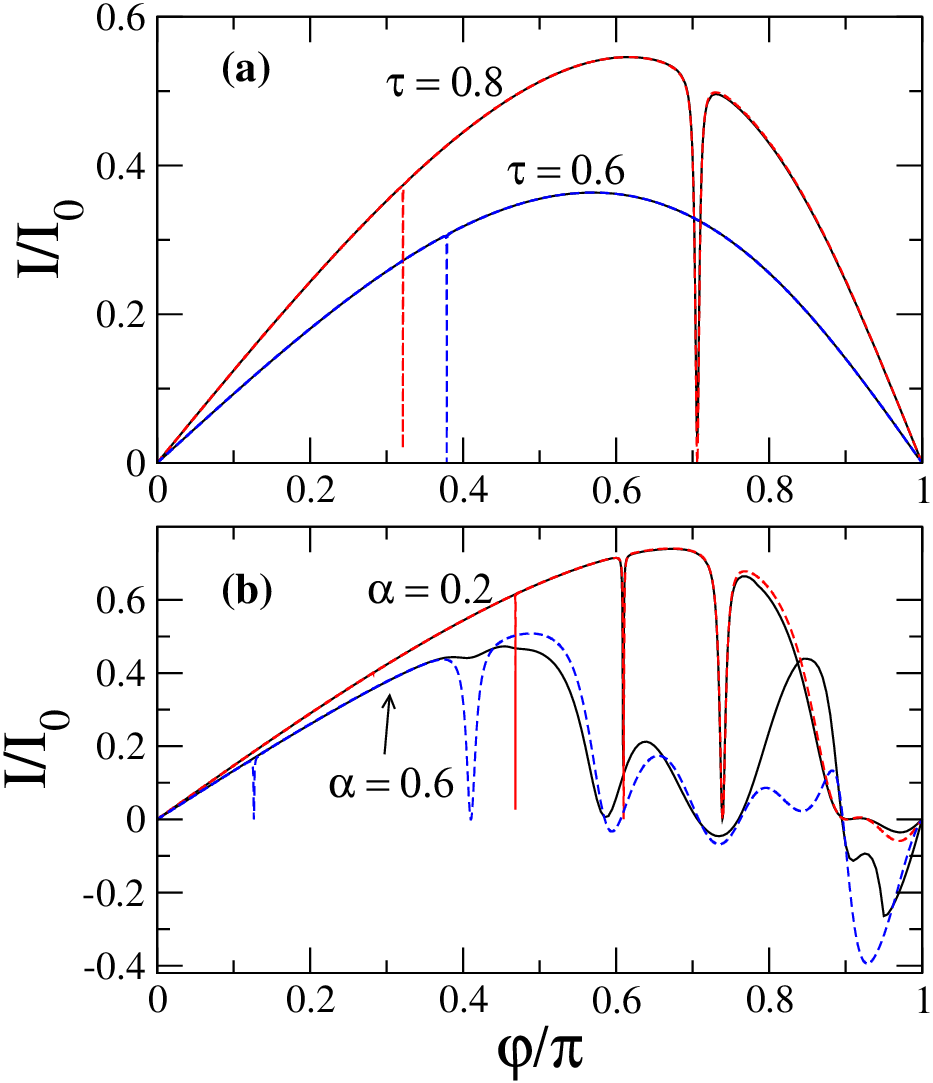}
\caption{\label{comparison2}
(Color online) (a) The current-phase relation for $\alpha=0.1$, $\hbar\omega=0.6\Delta$
and two values of the transmission coefficient, $\tau=0.8$ and $\tau=0.6$.
(b) The current-phase relation for $\hbar\omega=0.3\Delta$, $\tau=0.95$ and two values
of  $\alpha$, 0.2 and 0.6. In both panels the solid lines correspond to the microscopic
theory and the dashed lines to the two-level model.}
\end{center}
\end{figure}

The good agreement between the microscopic theory and the two-level model in these
examples can be  understood as follows. At zero temperature, the lower ABS is fully occupied,
while the upper one is empty. Therefore, for small values of $\alpha$ and $\hbar\omega<\Delta$
transfer of quasiparticles between the continuum and the ABSs is not possible. The agreement
between these models is further confirmed in Fig.~\ref{comparison2}(a), where the CPR
is shown for $\hbar\omega=0.6\Delta$, $\alpha=0.1$ and two lower values of the transmission
($\tau=0.6$ and $0.8$). In this case, the agreement is almost perfect for all phases.
The reason is that now the smallest energy gap between the ABSs, which occurs at
$\varphi=\pi$, is large enough to avoid the overlap of the levels in the presence of the
microwave field. If the transmission is further reduced, no transitions can occur
between the Andreev states and the adiabatic approximation becomes exact.

From the discussion above, we can conclude that the two-level model provides an
excellent description of the supercurrent at zero temperature and for weak fields
($\alpha \ll 1$). However, as the radiation power increases, the situation changes.
This is illustrated in Fig.~\ref{comparison2}(b), where we show the CPR for a highly
conductive channel ($\tau=0.95$), a frequency $\hbar \omega = 0.3\Delta$
and two values of $\alpha$. As one can see, the deviations between the
results of the two-level model and the microscopic theory become more apparent
as the power increases. The main reason for this discrepancy is the occurrence
of multiphotonic processes, which become more probable as the power increases.
These processes induce quasiparticle transitions between the ABSs and the continuum
part of the energy spectrum, which are not included in the two-level model.

\begin{figure}
\begin{center}
\includegraphics[width=0.85\columnwidth,clip]{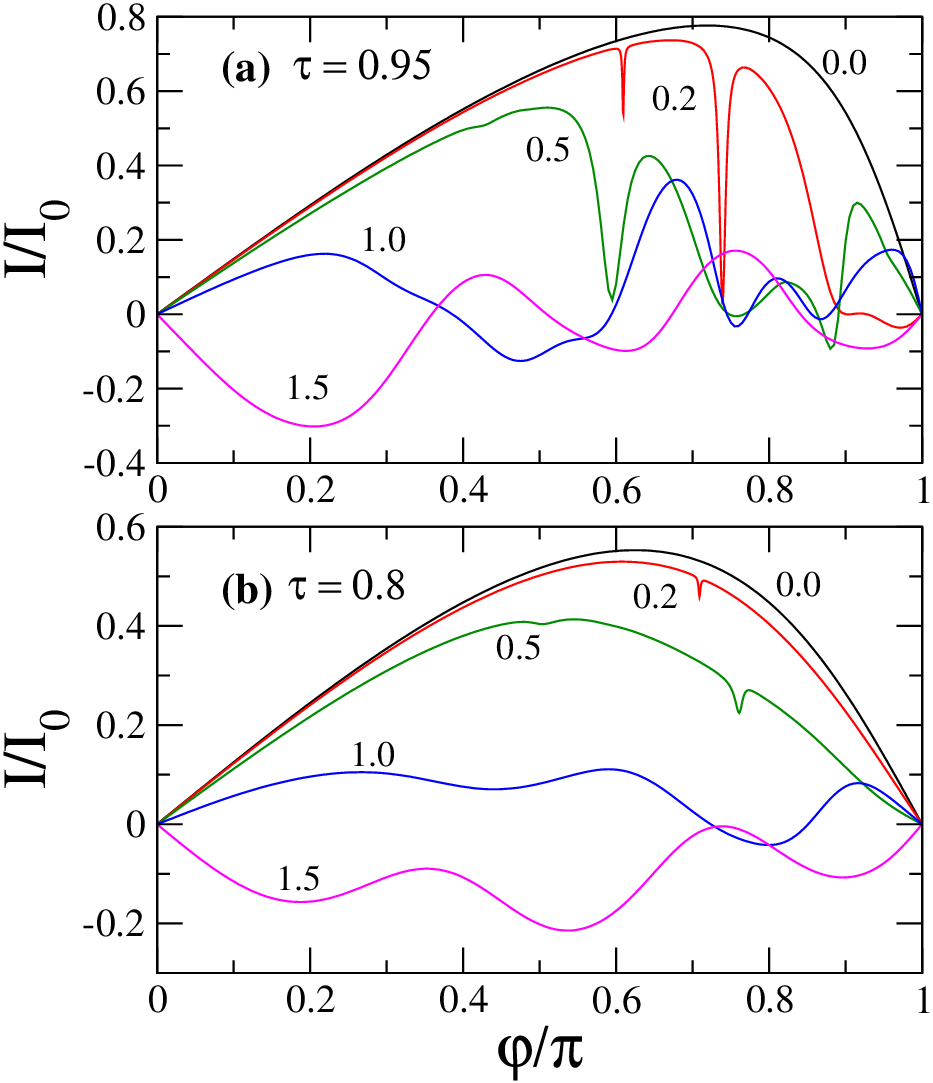}
\caption{\label{CPR-power} (Color online)
The zero-temperature current-phase relation for $\hbar\omega=0.3\Delta$
and two values of the transmission: (a) $\tau = 0.95$ and (b) $\tau = 0.8$.
The different curves correspond to different values of $\alpha$, as indicated
in the graphs. The inelastic broadening used in these calculations is $\eta =
10^{-3} \Delta$.}
\end{center}
\end{figure}

As one could already see in Fig.~\ref{comparison2}(b), as the radiation power
increases the supercurrent dips broaden and the CPR acquires a very rich
structure. We illustrate this fact in more detail in Fig.~\ref{CPR-power}
where we show the evolution of the CPR with $\alpha$ for two values of the
transmission (0.95 and 0.8) and for frequency $\hbar \omega = 0.3\Delta$. Notice
that as the power increases, the dips disappear, the CPRs become highly non-sinusoidal,
and in some regions of the phase the current can reverse its sign. These results
are clearly at variance with those found within the adiabatic approximation
(see Section II). They are a consequence of a complex interplay between the
dynamics of the ABSs, which are broadened by the coupling to the microwaves,
and the multiple transitions induced between the ABSs and the continuum of
states. This very rich behavior has also important implications for the critical
current, which for high transmission strongly deviates from the standard behavior
described by the adiabatic approximation. This is discussed below in detail.
Finally, it is worth stressing that the values of $\alpha$ used in
Fig.~\ref{CPR-power} are easily achievable in experiment, as demonstrated in the context
of atomic contacts,\cite{Chauvin2006} semiconductor nanowires\cite{Doh2005}
or graphene junctions.\cite{Heersche2007,Du2008,Jeong2011} Therefore, these results
indicate that the microscopic theory presented here will always be necessary for the description of  the experimental results of highly transmissive junctions at sufficiently
high power, no matter how low the microwave frequency is.

\subsection{Finite temperature: Enhancement of the supercurrent}\label{MMb}

We now turn to the analysis of the supercurrent at finite temperature, carried out within the microscopic  model. The new ingredient at
finite temperature is the fact that the ABSs are neither fully occupied
nor fully empty, which means that quasiparticle transitions between the
continuum of states and the bound states are possible, even for
frequencies $\hbar \omega < \Delta$. This has important
consequences.

\begin{figure}[t]
\includegraphics[width=0.8\columnwidth,clip]{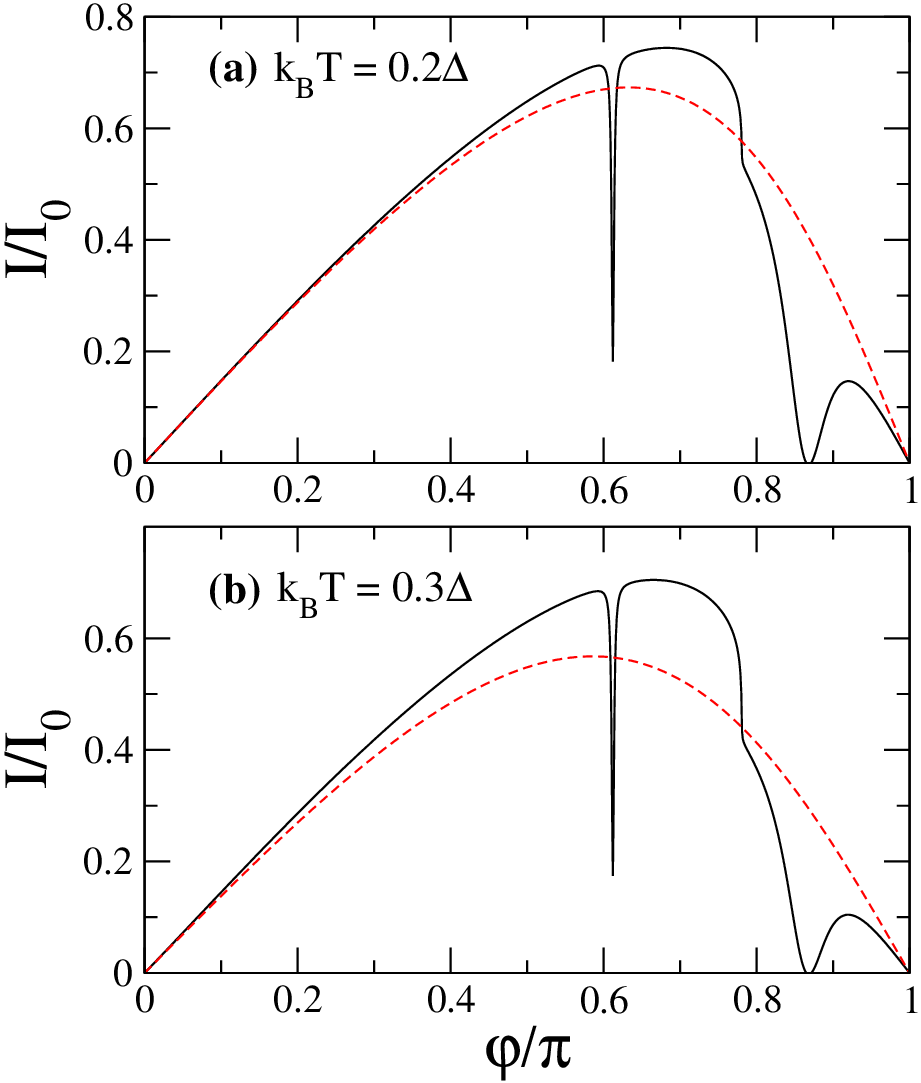}
\caption{\label{CPR-temp1}
(Color online) The current-phase relation for $\hbar \omega=0.6\Delta$, $\alpha=0.1$,
$\tau=0.95$ and two different temperatures: (a) $k_B T=0.2\Delta$ and (b) $k_B T=0.3\Delta$.
The solid lines in both panels correspond to the results of the microscopic theory and
the dashed lines to the supercurrent in the absence of microwaves ($\alpha=0$).}
\end{figure}

In Fig.~\ref{CPR-temp1} we show the CPR for $\hbar\omega=0.6\Delta$, $\alpha=0.1$,
$\tau=0.95$ and two different temperatures. For comparison, we also show the
results in the absence of microwaves (dashed lines). Apart from the dips, whose
origin is discussed above in detail, one can observe that at a certain value
of the phase ($\varphi_0 \approx 0.78\pi$) the current is suppressed. Notice
that the suppression is stronger as the temperature increases. Moreover, for phases
smaller than $\varphi_0$ the supercurrent exceeds its value in the absence of
microwaves. In other words, for $\varphi<\varphi_0$ there is an enhancement of
the supercurrent induced by the microwave field. The origin of this enhancement
is the promotion of quasiparticles from the continuum below $-\Delta$ to the lower
ABS by the microwave field.  There is also an identical  contribution coming from transitions connecting the upper state and the continuum
above $+\Delta$. At low microwave powers,  these processes can only occur if the field frequency is larger
than the distance in energy between the gap edges and the nearest ABS, i.e.,
if $\hbar \omega > \Delta-E_{\rm A}(\varphi)$, and they become possible at finite
temperature because the lower state is not fully occupied and the upper state
is not fully empty. For the parameters of Fig.~\ref{CPR-temp1} the previous
condition is satisfied if $E_{\rm A}(\varphi)>0.4\Delta$, which corresponds to a phase
$\varphi<0.78\pi$. Obviously, this phenomenon of microwave-enhanced supercurrent
cannot be described by the two-level model since this models ignores the contribution
of the continuum part of the spectrum.

In order to confirm our interpretation of the origin of the microwave-enhanced
supercurrent, we have derived analytical results describing this phenomenon
in the limit of weak microwave fields. We have obtained such results with the help
of an alternative method, known as Hamiltonian approach, which for SQPCs has been
shown to be equivalent to the microscopic theory described at the beginning of this
section.\cite{Cuevas1996,Cuevas2001,Bergeret2005} In this approach, a point
contact is described in terms of a tight-binding-like Hamiltonian and the transport
properties are calculated following a perturbative approach, where the coupling
between the electrodes is treated as the perturbation. Although the calculations with
this method are slightly more cumbersome than with the approach used above, it
has certain advantages. For instance, it also allows us to obtain the density of
states (DOS) at the contact. Moreover, a perturbative analysis (in the field) is
much simpler when using Eq.~\eqref{Idc}. The technical details of the Hamiltonian
approach are described in the Appendix A, and in what follows, we  only discuss
the results of this analysis.

\begin{figure}
\begin{center}
\includegraphics[width=0.8\columnwidth,clip]{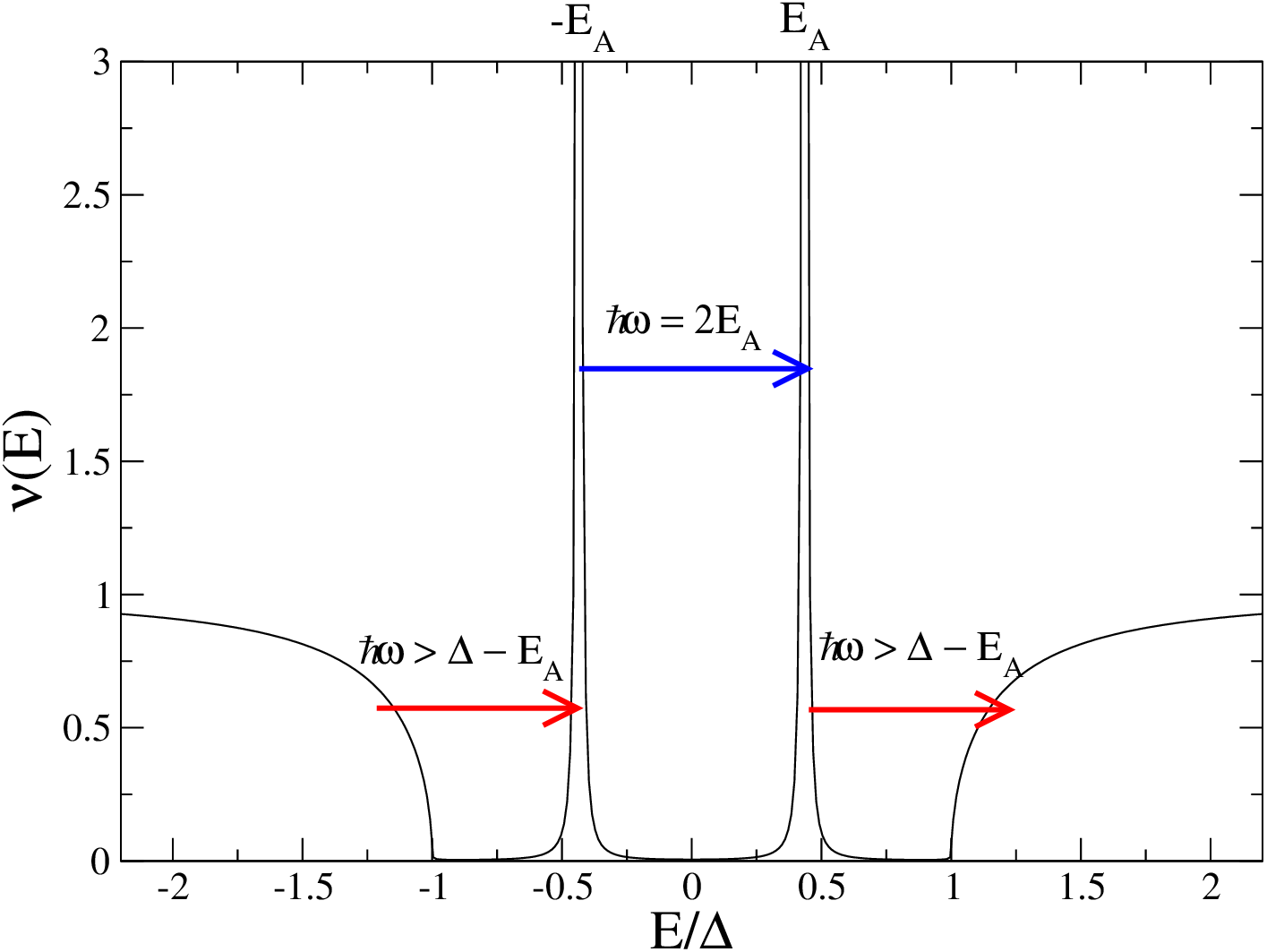}
\caption{\label{figdos}
(Color online) The local density of states at the contact in the absence of
microwaves, as defined in Eq.~\eqref{tbdos}, as a function of energy for
$\tau=0.95$, $\varphi=3\pi /4$ and $\eta=10^{-3}\Delta$. The lower
arrows represent the microwave-induced transitions between the continuum
part of the spectrum and the Andreev bound states which are responsible for the
supercurrent enhancement at finite temperatures. The upper arrow indicates the
resonant transition between the ABSs, which suppresses the supercurrent.}
\end{center}
\end{figure}

We are interested in the correction to the current due to the microwave field which
is responsible for the enhancement of the supercurrent. Thus, based on our numerical
results, we explore the parameter region where $\Delta-E_A < \hbar \omega$. Moreover, in order
to avoid the resonant transitions between the ABSs, we also assume that $\hbar \omega<2E_A$.
As described in Appendix A, a perturbative analysis to lowest order in the parameter
$\alpha$ shows that the supercurrent can be written as
\begin{equation}
I(\varphi) = I_{\rm eq}(\varphi) + \delta I(\varphi) , \label{anacurr}
\end{equation}
where $I_{\rm eq}$ is the equilibrium supercurrent given by Eq.~\eqref{Ieq},
and the correction $\delta I$ contains several contributions of order $\alpha^2$.
There are two types of contributions.
One type of  contribution is related to the change in the bound states induced by the
coupling to the electromagnetic field. The other contribution comes from the modification
of the occupations of the bound states due to the quasiparticle transitions involving
the ABSs. In the range of parameters that we are interested in, the second type of
contributions dominates at high enough temperatures and, in particular, they are
responsible for the supercurrent enhancement. Those contributions can be written in
the spirit of Eq.~\eqref{Ieq} as 
\begin{equation}
\label{eq-deltaI}
\delta I_{\rm enh}(\varphi) = I^-_{\rm A}(\varphi) \delta n^-(\varphi)
+ I^+_{\rm A}(\varphi) \delta n^+(\varphi) ,
\end{equation}
where  $I^{\pm}_{\rm A} (\varphi) = (2e/\hbar) \partial
E^{\pm}_{\rm A}/ \partial \varphi$ give the contribution of the states to the
equilibrium supercurrent, and $\delta n^{\pm}(\varphi)$ are the corrections to
the occupations of the ABSs due to the application of the microwave field. These
corrections can be written as
\begin{eqnarray}
\label{eq-deltan}
\delta n^{\pm}(\varphi) & = & \frac{\alpha^2 \tau}{8} \left[ {\rm Re} \left\{ e^{i\varphi}
\rho_L(E^{\pm}_{\rm A}) \tilde{\rho}_R(E^{\pm}_{\rm A} \pm \omega) \right\} + \right.
\nonumber \\
& & \left. \nu(E^{\pm}_{\rm A}) \nu(E^{\pm}_{\rm A} \pm \omega) \right] [F_0- F_{\pm 1}] .
\end{eqnarray}
Here, $F_n$ is the distribution function with shifted arguments $F_n =
\tanh[(E+n\hbar\omega)/2k_{\rm B}T]$, $\nu (E)$ is the density of states at the
contact in the absence of microwaves, and $\rho_j$ and $\tilde\rho_j$ are the
real part of the anomalous Green functions on the left (L) and (R) side of the
interface ($j=L,R$) without the  field, as defined in Appendix A.
Equation (\ref{eq-deltan}) has a very appealing form and it tells us that
the occupations of the ABSs can be changed by microwave-induced transitions
connecting these states between the continua below and above the gap. These
transitions are illustrated in Fig.~\ref{figdos}, where we also present an
example of the density of states of the contact in the absence of microwaves,
$\nu (E)$. This density of states is given by (see Appendix A)
\begin{equation}
\nu(E) = {\rm Re} \left\{ \frac{E \sqrt{(E+i\eta)^2-\Delta^2}}
{(E+i\eta)^2-E_{\rm A}^2} \right\}\; ,\label{tbdos}
\end{equation}
where the poles correspond to the ABSs and, as one can see in Fig.~\ref{figdos},
there are no  singularities at the gap edges $E=\pm \Delta$.

\begin{figure}
\begin{center}
\includegraphics[width=0.9\columnwidth,clip]{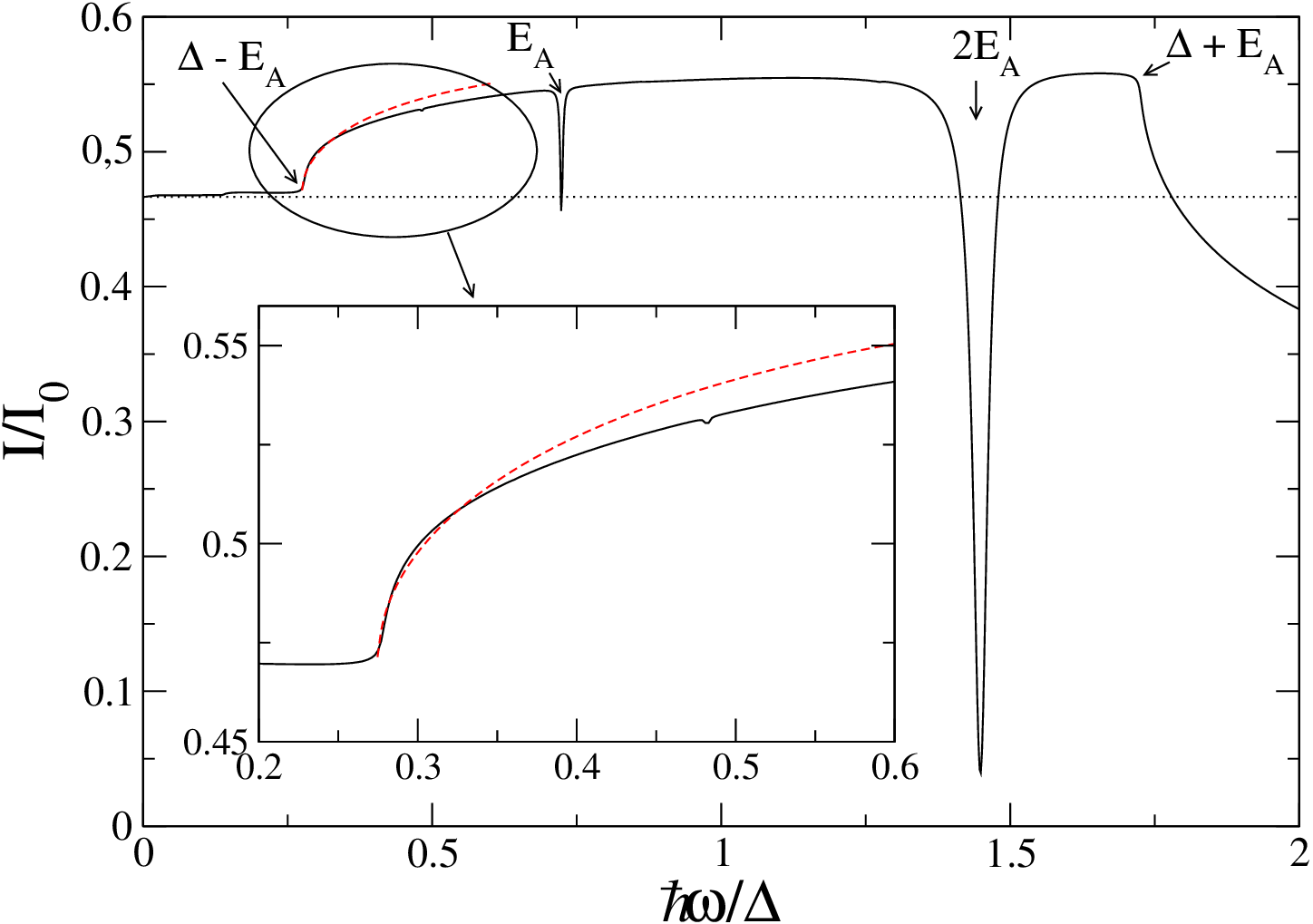}
\caption{\label{figwdp}
(Color online) The dc Josephson current as a function of the frequency $\omega$
of the microwave field for a fixed value of the phase $\varphi=\pi/2$, and
$\alpha=0.1$, $\tau=0.95$ and $k_BT=0.4\Delta$. The solid line shows the exact numerical result while the dashed line shows the result obtained from Eq. (\ref{corrcurr}). The dotted line shows the value of the current in the absence of the microwave field.}
\end{center}
\end{figure}

From Eq.~(\ref{eq-deltan}) one can show that the transitions between the
continuum of states below $-\Delta$ and the lower ABS increase the population
of the lower  state ($\delta n^{-} > 0$), while the photon processes connecting
the continuum above $+\Delta$ and the upper ABS decrease the
occupation of the upper state ($\delta n^{+} < 0$). As one can see from
Eq.~(\ref{eq-deltaI}), both types of processes give a positive contribution to
the current at finite temperatures and thus, they are responsible for the
supercurrent enhancement. Indeed, due to the electron-hole symmetry of this
problem, terms in Eq.~(\ref{eq-deltaI}) give the same contribution to the
current. Finally, the correction to the current due to these microwave-induced
transitions involving the continuum can be written as
\begin{eqnarray}
\delta I_{\rm enh}(\varphi) &  = & \alpha^2 \left( \frac{-2e E^{\prime}_{\rm A}}{\hbar}
\right) \frac{\tau}{16} \times \label{corrcurr} \\
& & \frac{\sqrt{(E_{\rm A}+\hbar\omega)^2 - \Delta^2} \sqrt{\Delta^2-E_{\rm A}^2}}
{\eta \hbar \omega E_{\rm A}(2E_{\rm A}+\hbar\omega) } \times \nonumber \\
& & \hspace*{-1.6cm} \left[ E_{\rm A} \hbar \omega + \Delta^2 ( 1+ \cos \varphi) \right]
\left[ F_1- F_0 \right] \Theta\left(|E_{\rm A}+\hbar\omega |-\Delta\right).
\nonumber
\end{eqnarray}
This expression gives a positive contribution to the supercurrent and it explicitly
shows that the enhancement can only take place when $\hbar \omega > \Delta-E_{\rm A}$. 
{ According to Eq. (\ref{corrcurr}),  the correction to the current  is proportional $1/\eta$, the inelastic scattering time. In our model the parameter $\eta$ describes the energy loss mechanism via which the microwave power is dissipated.  For simplicity, we assume it to be energy and frequency independent.  }
Equation (\ref{corrcurr})  reproduces the exact results obtained with the microscopic approach in the limit of weak fields and in the range of frequencies where the transitions
between the ABSs cannot take place. This is illustrated in Fig.~\ref{figwdp} where we
show the supercurrent for a fixed value of the phase ($\varphi = \pi /2$) as
a function of the frequency for $\tau = 0.95$, $\alpha = 0.1$ and $k_{\rm B}T =
0.4\Delta$. As one can see, the exact result (solid line) remains constant for low
frequencies. Then, at $\hbar \omega = \Delta - E_{\rm A}$ there is a rise of the
supercurrent due to the onset of the transitions connecting the ABSs with the
continuum of states. This increase of the supercurrent is well described by the analytical
result of Eq.~(\ref{corrcurr}) (dashed line). At higher frequencies, one can
observe the dips due to the transitions between the ABSs. The dip at $\hbar \omega =
E_{\rm A}$ corresponds to a two-photon process, while the one at $\hbar \omega =
2E_{\rm A}$ is produced by a single-photon process. Finally, at $\hbar \omega =
\Delta + E_{\rm A}$ the supercurrent starts to decrease due to microwave-induced
transitions between the continuum below $-\Delta$ and the upper ABS and similar
ones between the continuum above $+\Delta$ and the lower ABS. These transitions,
which can also occur at zero temperature, tend to increase the occupation of the
upper state and to reduce the population of the lower one, which results in a
reduction of the net supercurrent.

As one can see in Fig.~\ref{CPR-temp1} and ~\ref{figwdp}, the maximum supercurrent
sustained by the junction, i.e.\ the critical current, can also be enhanced by the
microwave field at finite temperatures. A microwave-enhanced critical current was
first reported in experiments on superconducting microbridges\cite{Wyatt1966,Dayem1967}
and explained by Eliashberg\cite{Eliashberg1970} in 1970 in terms of the stimulation
of the superconductivity in the electrodes, which were made of thin films. Such
a stimulation, and the corresponding microwave-enhanced critical current, only
occur at temperatures very close to the critical temperature. Enhancements at
much lower temperatures were reported in the 1970's in the context of SNS
structures,\cite{Notarys1973,Warlaumont1979} and they have been recently explained
in terms of the redistribution of the quasiparticles induced by the field.\cite{Virtanen2010}
In this case, for the enhancement to occur, the temperature must be of the order
of the minigap in the normal wire, which can be much lower than the critical
temperature of the superconducting leads.

\begin{figure}
\begin{center}
\includegraphics[width=0.9\columnwidth,clip]{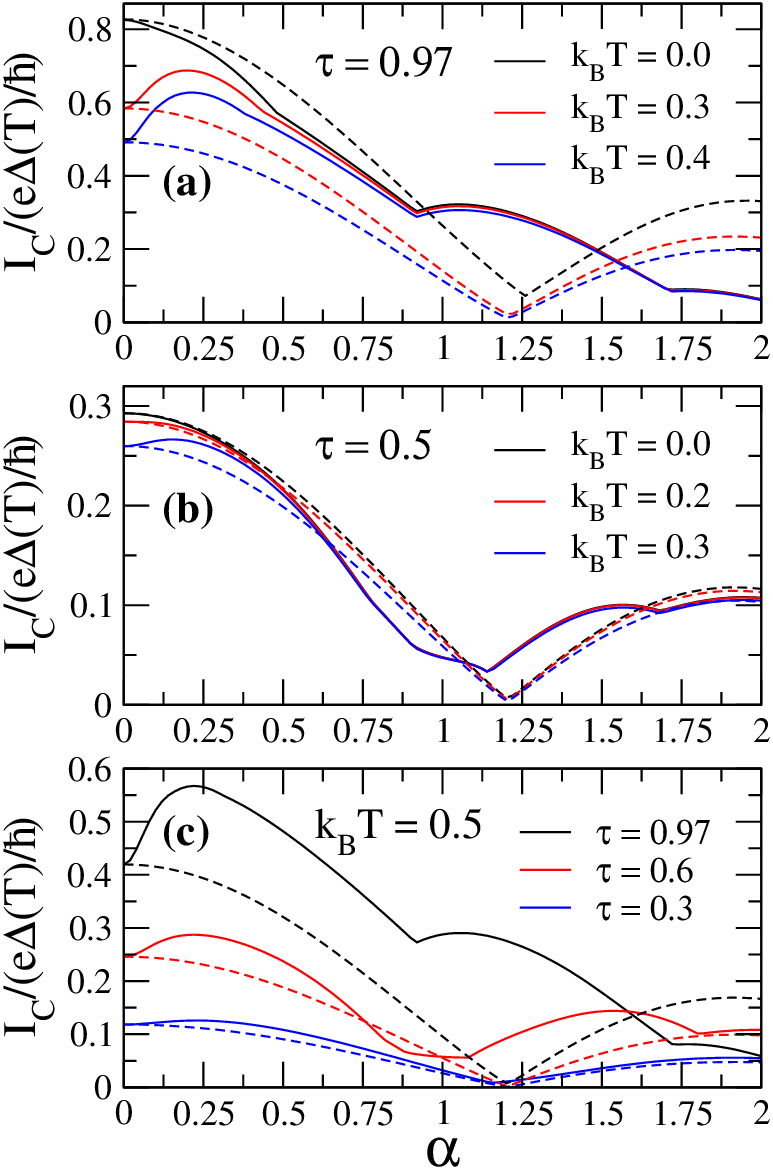}
\caption{\label{Icritical}
(Color online) The critical current as a function of $\alpha$ for $\hbar \omega =
0.6\Delta$. The different curves correspond to different values of the temperature
and the transmission as indicated in the graphs. The solid lines correspond to
the exact results, while the dashed lines show the results of the adiabatic
approximation. In the three panels the critical current has been normalized by
$e\Delta(T)/ \hbar$, where $\Delta(T)$ is the gap at the corresponding temperature.}
\end{center}
\end{figure}

As discussed above, in the case of a point contact the mechanism is similar to
that of diffusive SNS structures,\cite{Virtanen2010} but it involves discrete
ABSs, rather than a continuous band of ABSs, as in the case of
diffusive proximity structures. For this reason we may expect  the enhancement of the critical
current in SQPCs  to occur at intermediate temperatures, when $k_{\rm B}T$
is of the order of the energy distance between the ABSs and the gap edges ($\Delta
- E_{\rm A}(\varphi_{\rm max})$), where $\varphi_{\rm max}$ is the phase value
at which the supercurrent reaches its maximum. This is illustrated in
Fig.~\ref{Icritical}, where we show the critical current as a function of $\alpha$
for different temperatures and different  values of the transmission. 
Panel (a) shows the critical current for a highly transmissive channel ($\tau =
0.97$) and three values of the temperature. Notice first that at finite temperatures,
the critical current at finite $\alpha$ ($\alpha \lesssim 0.5$) exceeds the value
in the absence of microwaves ($\alpha=0$). Notice also that as $\alpha$ increases,
the critical currents clearly deviate from the behavior described by the adiabatic
approximation, which is shown  as dashed lines.
It is also important to emphasize that the microwave-enhancement of the critical current
is not exclusive of high conductive channels and it persists up to relatively low
transmissions, as we show in Fig.~\ref{Icritical}(b-c).  The relative enhancement of the critical current is larger the larger is the temperature. 
It is also worth remarking that at sufficiently high power, the critical current depends only weakly on the temperature.

\section{Conclusions and further discussions}

Summarizing, we have presented a theoretical analysis of the supercurrent in a
phase-biased SQPC under microwave irradiation. We have shown that if the microwave
frequency $\omega$ is not high enough to induce transitions between the ABSs or
between the ABSs and the continuum of states outside the gap region, the supercurrent
is correctly described by the standard adiabatic approximation (see Section II).
However, when $\hbar \omega$ is comparable to the Andreev gap (energy distance between
the ABSs),  quasiparticle transitions between the ABSs can occur and the
supercurrent can be largely suppressed at the corresponding values of the phase
difference. We have shown that this phenomenon can be nicely explained within a
two-level model that describes the dynamics of the ABSs.\cite{Zazunov2003} This
model indicates that the supercurrent suppression is due to the enhancement of
the occupation of the upper ABS induced by resonant transitions from the lower state.
Moreover, at low temperatures and weak fields, this model is quantitatively correct
provided that (i) the microwave frequency is not high enough to induce transitions
connecting the ABSs and the continuum of states, and (ii) the Andreev gap is large
compared to the broadening acquired by the ABSs by means of the coupling to the
electromagnetic field. Finally, we have shown that whenever microwave-induced
transitions between the ABSs and the continuum of states become possible (due to
finite temperatures, high frequencies or high radiation powers), a fully microscopic
theory is needed to describe the supercurrent. We have developed such a theory and
predicted the following effects. First, at finite temperatures it is possible to
enhance both the supercurrent and the critical current by the application of a
microwave field. This effect originates from the quasiparticle transitions between
the ABSs and the continuum of states, which increase the occupation of the lower
Andreev state and reduce the population of the upper one. Second, the current-phase
relation at high powers is strongly distorted and it can become highly non-sinusoidal
exhibiting sign changes in the region between $0$ and $\pi$. Third, the critical current
as a function of the radiation power can exhibit large deviations from the standard
Bessel-function behavior described by the adiabatic approximation.

It is now pertinent to discuss the connection with experiments. As explained in
the introduction, most of the experimental results of the effect of microwaves
on the supercurrent of a point contact have been successfully described in the frame
of the adiabatic approximation. The reason is that the typical frequency used in
the experiments is relatively low ($\hbar \omega \ll \Delta$) and no transitions
between the ABSs can occur. However, it is important to remark that there are no
fundamental limitations to study the parameter regime where we predict the occurrence
of novel effects like the appearance of supercurrent dips in the current-phase
relation or the microwave-enhanced critical current. These effects are easier to
observe in highly transmissive point contacts where the Andreev gap can become
relatively small (much smaller than $\Delta$). The ideal experimental system
where to test our predictions is a superconducting atomic contact for several
reasons. First, these contacts can sustain a reduced number of channels, which
facilitates the comparison with the theory. Second, it has been shown that it
is possible to determine independently the set of transmission eigenvalues $\{
\tau_i \}$,\cite{Scheer1997} which has allowed to establish a comparison between
theory and experiment with no adjustable parameters for many different transport
properties.\cite{Cron2001,Chauvin2006,Rocca2007}. Third, it is possible to tune,
at least to a certain extent, the transmission coefficients and, in particular,
to achieve very high transmission coefficients, as demonstrated in the context of Al
atomic contacts.\cite{Scheer1997,Chauvin2006,Rocca2007} Finally, it has already
been shown that in these systems the current-phase relation is amenable to
measurements,\cite{Rocca2007} and investigations of the transport properties
of superconducting atomic contacts under microwave irradiation have already
been performed.\cite{Chauvin2006,note-exp}

\begin{figure}
\begin{center}
\includegraphics[width=0.9\columnwidth,clip]{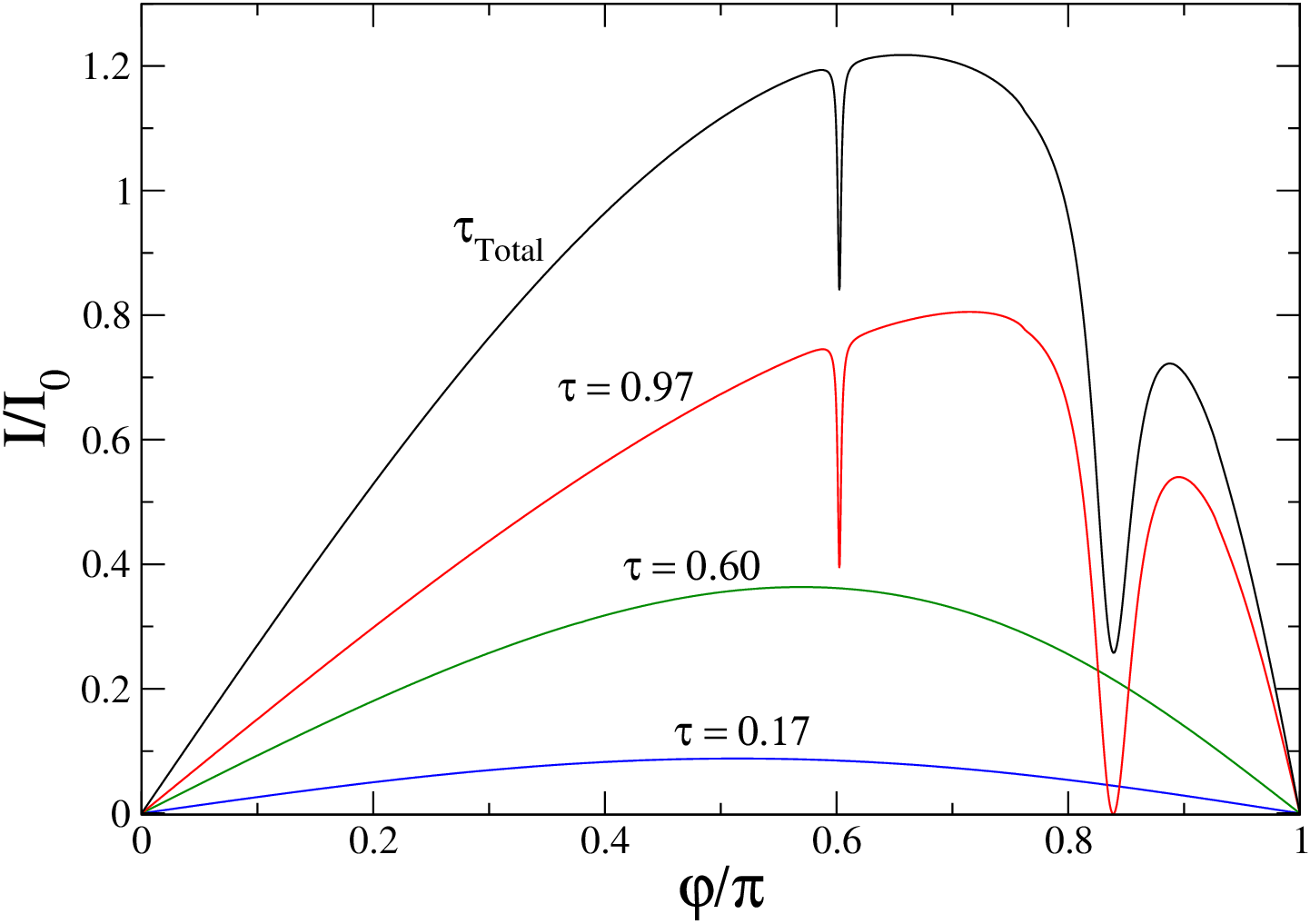}
\caption{\label{3chan}
(Color online) The zero-temperature current-phase relation for a point-contact consisting of three channels with transmissions $\tau=0.17,0.6,0.97$, for $\alpha=0.1$ and $\hbar\omega=0.6\Delta$. The dashed lines show the contribution of each channel, while the solid line corresponds to the  total current.}
\end{center}
\end{figure}

In  experiments with superconducting atomic junctions, even at the level of a
single-atom contact, one often has the contribution of several conduction
channels. In this sense, one may wonder whether the presence of low-transmissive
channels can mask some of the striking effects that we have discussed above.
In Fig. \ref{3chan} we show the CPR for a contact consisting of three conducting channels with transmissions $\tau=0.17,0.6,0.97$ respectively.  The current is obtained by adding the contribution of  each  channel according to  Eq. (\ref{matrixJ}). As one can see in  Fig. \ref{3chan} the total current  still shows the dips at the resonances corresponding  to the channel with the highest transmission ($\tau=0.97$).  However, the current does not vanish completely   due to the contribution of the  low-transmissive channels.

It is worth stressing that the major problem to establish a direct comparison
between our theory and the experiments is the fact that we have assumed a
phase-biased junction. In reality, and depending on the details of the
electromagnetic environment seen by the point contact, the phase across the
junction may undergo fluctuations (both classical and quantum) which can
affect the value of the critical current or the shape of the current-phase
relation. Thus, a quantitative comparison with the experiments may require
in some cases to combine our theory with a description of the phase
fluctuations. For classical fluctuations, this could be done in the spirit of
Ref.~\onlinecite{Chauvin2007} by means of an extension of the resistively shunted
junction using our microscopic current-phase relation as a starting point.

Let us conclude by saying that in this work we have shown that the application of
microwaves to one of the simplest superconducting systems, namely a SQPC, leads to
a very rich phenomenology, which has remained largely unexplored. In particular, we
have shown that a microwave field is an ideal tool to make a direct spectroscopy
of the Andreev bound states of a superconducting junction. The ideas put
forward in this work pave the way for the understanding of the influence of
a microwave field on the supercurrent of variety of highly transmissive
superconducting weak links.

\begin{acknowledgments}
We thank A. Levy Yeyati, C. Urbina, M. Feigelman, M. A. Cazalilla, R. Avriller and C. Tejedor
for discussions that in part motivated this work. This work was supported by the
Academy of Finland, the ERC (Grant No.\ 240362-Heattronics), the Spanish MICINN
(Contract No.\ FIS2008-04209), the Emmy-Noether program of the DFG, and the
CSIC (Intramural Project No.\ 200960I036).
\end{acknowledgments}

\appendix

\section{Hamiltonian approach}

The transport properties of a microwave-irradiated SQPC can also be described
within the so-called Hamiltonian approach.\cite{Cuevas1996,Cuevas2002} We 
explain in this appendix how this approach can be used to obtain analytical
results for the supercurrent enhancement discussed in Section \ref{MMb}. In this
approach a single-channel SQPC can be described in terms of the following
tight-binding-like Hamiltonian
\begin{equation}
\hat H = \hat H_L + \hat H_R + \sum_{\sigma} \left\{ t \hat c^\dagger_{L \sigma}
\hat c_{R \sigma} + t^{\ast} c^\dagger_{R \sigma} \hat c_{L \sigma} \right\},\label{tbh}
\end{equation}
where $\hat H_{L,R}$ are the BCS Hamiltonians describing the left (L) and
right (R) electrodes and the last term describes the coupling between the
electrodes. In this last term, $t$ is a hopping element that determines
the transmission of the contact.

In this model the current evaluated at the interface between the two electrodes
adopts the form
\begin{equation}
I(t) = \frac{ie}{\hbar} \sum_{\sigma} \left\{ t \langle c^\dagger_{L \sigma}
\hat c_{R \sigma} \rangle - t^{\ast} \langle c^\dagger_{R \sigma} \hat c_{L \sigma}
\rangle \right\} .
\end{equation}
This expression can be rewritten in terms of the Keldysh Green functions as

\begin{equation}
I(t) = \frac{e}{\hbar}{\rm Tr}\bigl [\hat \tau_3\bigl(
\hat t \hat{\cal G}_{RL}^K -\hat t^\dagger \hat{\cal G}_{L R}^K\bigr)\bigr](t,t)
\;. \label{tbc}
\end{equation}
Here $\hat \tau_3$ is the corresponding Pauli matrix, Tr denotes the trace
in Nambu space, and $\hat t$ is the hopping matrix in Nambu space given by
\begin{equation}
\label{hopping-matrix}
\hat t = \left( \begin{array}{cc}
t e^{i \phi(t)/2} & 0 \\
0 & -t^{\ast} e^{-i \phi(t)/2} \end{array} \right) .
\end{equation}
Here, $\phi(t)$ is the time-dependent superconducting phase given by
Eq.~(\ref{eq-phi}).

In order to determine the Green functions appearing in the current expression,
we follow a perturbative scheme and treat the coupling term in Hamiltonian
(\ref{tbh}) as a perturbation. The unperturbed Green functions describe the
uncoupled electrodes in equilibrium. Thus for instance, the retarded and
advanced functions are given by
\begin{equation}
\hat g_{jj}^{R(A)}(E) = \frac{-i}{W} \frac{1}{\zeta^{R(A)}(E)}
\begin{pmatrix} E & \Delta\\ \Delta & E
\end{pmatrix} \; ,
\end{equation}
where $j=L,R$, $\zeta^{R(A)}=\sqrt{(E+i\eta)^2-\Delta^2}$, and $W$ is an energy
scale related to the normal density of states at the Fermi energy. The full
Green functions can then be determined by solving a Dyson equation, where
the retarded and advanced self-energies are simply given by the hopping
matrix of Eq.~(\ref{hopping-matrix}).

Since we are interested in the limit of weak fields ($\alpha \ll 1$), we
can expand the phase factors in Eq.~(\ref{hopping-matrix}) as follows
\begin{equation}
e^{i \phi(t)/2} \approx e^{i \varphi/2} \left( 1 + \alpha \cos \omega t +
\frac{1}{2} \alpha^2 (\cos \omega t)^2 + \cdots \right) .
\end{equation}
Moreover, for the perturbative treatment in $\alpha$ it is convenient to use
the full Green functions of the contact in the absence of microwaves ($\alpha=0$),
$\hat G_{ij}$. It is straightforward to show that these functions can be
expressed as
\begin{equation}
\hat G_{LL}^{R(A)} = \frac{-i\zeta^{R(A)}}{W(1+\beta)\xi^{R(A)}}
\begin{pmatrix}E\pm i\eta & E_g^*\\E_g  &E \pm i\eta \end{pmatrix}\;\label{GLL}
\end{equation}
\begin{equation}
\hat G_{RL}^{R(A)} = \frac{-t}{W^2(1+\beta)\xi^{R(A)}}
\begin{pmatrix}a^{R(A)} & b^{R(A)}\\-b^{R(A)*} & -a^{R(A)*} \end{pmatrix}\;,
\end{equation}
where $E_g=\Delta(1+\beta e^{i\varphi})(1+\beta)$, $\beta=(t/W)^2$, $\xi=E^2-E_A^2$,
$a=E^2e^{-i\varphi/2}-\Delta E_g^*e^{i\varphi/2}$ and $b=E(E_ge^{-i\varphi/2}-\Delta
e^{i\varphi/2})$. Similar expressions hold for $G_{RR}$ and $G_{LR}$. These Green
functions are now the zero-order propagators of the perturbation theory.  Substituting these functions in the current
expression of Eq.~(\ref{tbc}) and identifying the transmission coefficient as
$\tau=4\beta/(1+\beta)^2$,\cite{Cuevas1996} one obtains the expression
for the equilibrium current of Eq.~(\ref{Ieq}). On the other hand, from the
previous expressions one can determine the local density of states at the
contact  in the absence of microwaves, which is defined as $\nu_j(E) =
W(1+\beta)(i/2)(\hat G_{jj}^R-\hat G_{jj}^A)_{11}$ ($\nu_L = \nu_R$ in our
symmetric contacts). This density of states is given by Eq.~(\ref{tbdos})
and it is shown in Fig.~\ref{figdos}.

Going into the energy representation as  done  in Section \ref{MM}, the
first correction to the current of Eq.~(\ref{tbc}), which is of order $\alpha^2$,
contains the following three terms
\begin{equation}
\delta I = \frac{e}{\hbar} \int \frac{dE}{2\pi}{\rm Tr}\bigl[\hat t^{(2)}\hat{\cal G}_{RL}^{(0)} +
\hat t^{(0)}\hat{\cal G}_{RL}^{(2)}+\hat t^{(1)}\hat{\cal G}_{RL}^{(1)}\bigr]-
L\leftrightarrow R  \label{pert_c}\,,
\end{equation}
where the superindices denote the order of perturbation in $\alpha$. To obtain a
complete analytical expression for an arbitrary value of the field frequency is
quite cumbersome. Instead, we concentrate on the parameter range where the current
enhancement takes place. For that purpose, we focus on frequency values far from
the resonant condition $\hbar \omega=2E_{\rm A}$ and close to $\Delta-E_{\rm A}$.
In this region, it turns out that the second term in Eq.~(\ref{pert_c}) is
proportional to the parameter $\Delta/\eta$. All the other terms give a contribution
which depends only weakly on the frequency. Assuming a small inelastic scattering
rate, one can approximate the correction to the current by Eqs.~(\ref{eq-deltaI})
and (\ref{eq-deltan}), with  $\rho_j=(i/2)W(1+\beta) [\hat G_j^R
-\hat G_j^A]_{1,2}$, $\tilde\rho_j=(i/2)W(1+\beta)[\hat G_j^R-\hat G_j^A]_{2,1}$ ,
and $F_n=F(E+n\hbar\omega)$. This correction gives precisely the enhancement of
the supercurrent at finite temperatures, as discussed in section \ref{MMb}.

%%%%%%%%%%%%%%%%%%%%%%%%%%%%%%%%%%%%


\begin{thebibliography}{10}

\bibitem{Josephson1962}
B. D. Josephson, Phys. Lett. {\bf 1}, 251 (1962).

\bibitem{Anderson1963}
P. W. Anderson and J.M. Rowell, Phys. Rev. Lett. {\bf 10}, 230 (1963).

\bibitem{Likharev1979}
For a review on the activities in the 1960's and 1970's see
K. K. Likharev, Rev. Mod. Phys. {\bf 51}, 101 (1979) and Ref.\onlinecite{Barone1982}.

\bibitem{Barone1982}
 A. Barone and G. Paterno, {\it Physics and Applications of the
Josephson Effect} (Wiley-Interscience, New York, 1982).

\bibitem{Golubov2004}
For a recent review see A. A. Golubov, M. Yu. Kupriyanov, and E. Il'ichev,
Rev. Mod. Phys. {\bf 76}, 411 (2004).

\bibitem{Koops1996}
M. C. Koops, G. V. van Duyneveldt, and R. de Bruyn Ouboter,
Phys. Rev. Lett. {\bf 77}, 2542 (1996).

\bibitem{Goffman2000}
M. F. Goffman, R. Cron, A. Levy Yeyati, P. Joyez, M. H. Devoret, D. Esteve, and C. Urbina,
Phys. Rev. Lett. {\bf 85}, 170 (2000).

\bibitem{Rocca2007}
M. L. Della Rocca, M. Chauvin, B. Huard, H. Pothier, D. Esteve, and C. Urbina,
Phys. Rev. Lett. {\bf 99}, 127005 (2007).

\bibitem{Kasumov1999}
A. Yu. Kasumov, R. Deblock, M. Kociak, B. Reulet, H. Bouchiat, I. I. Khodos,
Yu. B. Gorbatov, V. T. Volkov, C. Journet and M. Burghard,
Science {\bf 284}, 1508 (1999).

\bibitem{Jarillo-Herrero2006}
P. Jarillo-Herrero, J. A. van Dam, and L. P Kouwenhoven,
Nature {\bf 439}, 953 (2006).

\bibitem{Jorgensen2006}
H. I. J{\o}rgensen, K. Grove-Rasmussen, T. Novotn{\'y}, K. Flensberg, and P. E. Lindelof,
Phys. Rev. Lett. {\bf 96}, 207003 (2006).

\bibitem{Kasumov2005}
A. Yu. Kasumov, K. Tsukagoshi, M. Kawamura, T. Kobayashi, Y. Aoyagi, K. Senba,
T. Kodama, H. Nishikawa, I. Ikemoto, K. Kikuchi, V. T. Volkov, Yu. A. Kasumov, R.
Deblock, S. Gu\'eron, and H. Bouchiat, Phys. Rev. B {\bf 72}, 033414 (2005).

\bibitem{Doh2005}
Y.-J. Doh, J. A. van Dam, A. L. Roest, E. P. A. M. Bakkers, L. P. Kouwenhoven and
S. De Franceschi, Science {\bf 309}, 272 (2005).

\bibitem{Xiang2006}
J. Xiang, A. Vidan, M. Tinkham, R. M. Westervelt, and C. M. Lieber,
Nature Nanotech. {\bf 1}, 208 (2006).

\bibitem{Heersche2007}
H. B. Heersche, P. Jarillo-Herrero, J. B. Oostinga1, L. M. K. Vandersypen, and
A. F. Morpurgo, Nature (London) {\bf 446}, 56 (2007).

\bibitem{Du2008}
X. Du, I. Skachko and E. Y. Andrei, Phys. Rev. B {\bf 77}, 184507 (2008).

\bibitem{Jeong2011}
D. Jeong, J.-H. Choi, G.-H. Lee, S. Jo, Y.-J. Doh, and H.-J. Lee,
Phys. Rev. B {\bf 83}, 094503 (2011).

\bibitem{Furusaki1991}
A. Furusaki and M. Tsukada, Solid State Commun. {\bf 78}, 299 (1991).

\bibitem{Beenakker1991}
C. W. J. Beenakker, Phys. Rev. Lett. {\bf 67}, 3836 (1991).

\bibitem{Desposito2001}
M. A. Desp\'osito and A. Levy Yeyati, Phys. Rev. B {\bf 64}, 140511 (2001).

\bibitem{Zazunov2003}
A. Zazunov, V. S. Shumeiko, E. N. Bratus', J. Lantz, and G. Wendin,
Phys. Rev. Lett. {\bf 90}, 087003 (2003).

\bibitem{Zazunov2005}
A. Zazunov, V. S. Shumeiko, G. Wendin, and E. N. Bratus',
Phys. Rev. B {\bf 71}, 214505 (2005).

\bibitem{Shumeiko1993}
V. S. Shumeiko, G. Wendin and E. N. Bratus', Phys. Rev. B {\bf 48}, 13129 (1993).

\bibitem{Gorelik1995}
L.Y. Gorelik, V.S. Shumeiko, R.I. Shekhter, G. Wendin, and M. Jonson,
Phys. Rev. Lett. {\bf 75}, 1162 (1995).

\bibitem{Gorelik1998}
L.Y. Gorelik, N.I. Lundin, V.S. Shumeiko, R.I. Shekhter, and M. Jonson,
Phys. Rev. Lett. {\bf 81}, 2538 (1998).

\bibitem{Lundin2000}
N. I. Lundin, Phys. Rev. B {\bf 61}, 9101 (2000).

\bibitem{Bergeret2010}
F. S. Bergeret, P. Virtanen, T. T. Hekkil\"a, and J.C. Cuevas,
Phys. Rev. Lett. {\bf 105}, 117001 (2010) .

\bibitem{Ivanov1999}
D.A. Ivanov and M.V. Feigelman, Phys. Rev. B {\bf 59}, 8444 (1999).

\bibitem{Haberkorn1978}
W. Haberkorn, H. Knauer, and J. Richter, Phys. Status Solidi A {\bf 47}, K161 (1978).

\bibitem{Ambegaokar1963}
V. Ambegaokar and A. Baratoff, Phys. Rev. Lett. {\bf 10}, 486 (1963).

\bibitem{Kulik1977}
I. O.  Kulik, and A. N. Omelyanchuk, Sov. J. Low Temp. Phys. {\bf 3}, 459 (1977).

\bibitem{Grifoni1998}
For a review on techniques and approaches for problems on driven tunneling
see M. Grifoni and P. H\"{a}nggi, Phys. Reps. {\bf 304}, 229 (1998) and references
therein.

\bibitem{Bloch1940}
F. Bloch and A. Siegert, Phys. Rev. {\bf 57}, 522 (1940).

\bibitem{Autler1955}
S. H. Autler and C. H. Townes, Phys. Rev. {\bf 100}, 703 (1955).

\bibitem{note-IC}
In all the numerical calculations within the two-level model shown in this work
we have assumed that the system is initially in its ground state, i.e.,
$\Psi_0^\dagger=(0,1)$.

\bibitem{Magnus54}
W. Magnus, Commun. Pure Appl. Math. {\bf 7}, 649 (1954).

\bibitem{Zaitsev1998}
A.V. Zaitsev and D.V. Averin, Phys. Rev. Lett. {\bf 80}, 3602 (1998).

\bibitem{Nazarov1999}
Yu. V. Nazarov, Superlattices Microstruct. {\bf 25}, 1221 (1999).

\bibitem{Cuevas2001}
J. C. Cuevas and M. Fogelstr\"om, Phys Rev B. {\bf 64}, 104502 (2001).

\bibitem{Kopu2004}
J. Kopu, M. Eschrig, J. C. Cuevas, and M. Fogelstr\"om,
Phys. Rev. B {\bf 69}, 094501 (2004).

\bibitem{Chauvin2006}
M. Chauvin, P. vom Stein, H. Pothier, P. Joyez, M. E. Huber, D. Esteve,
and C. Urbina, Phys. Rev. Lett. {\bf 97}, 067006 (2006).

\bibitem{Cuevas1996}
J. C. Cuevas, A. Martin-Rodero, and A. Levy Yeyati,
Phys Rev B. {\bf 54}, 7366 (1996).

\bibitem{Bergeret2005}
F. S. Bergeret, A. Levy Yeyati, and A. Mart\'{\i}n-Rodero,
Phys. Rev. B {\bf 72}, 064524 (2005).

\bibitem{Wyatt1966}
A.F.G. Wyatt, V. M. Dmitriev, W. S. Moore, and F. W. Sheard,
Phys. Rev. Lett. {\bf 16}, 1166 (1966).

\bibitem{Dayem1967}
A.H. Dayem and J.J. Wiegand, Phys. Rev. {\bf 155}, 419 (1967).

\bibitem{Eliashberg1970}
G. M. Eliashberg, JETP Lett. {\bf 11}, 114 (1970).

\bibitem{Notarys1973}
H. A. Notarys, M. L. Yu, and J. E. Mercereau,
Phys. Rev. Lett. {\bf 30}, 743 (1979).

\bibitem{Warlaumont1979}
J. M. Warlaumont, J. C. Brown, T. Foxe, and R. A. Buhrman,
Phys. Rev. Lett. {\bf 43}, 169 (1979).

\bibitem{Virtanen2010}
P. Virtanen, T. T. Heikkil\"a, F. S. Bergeret, and J. C. Cuevas,
Phys. Rev. Lett. {\bf 104}, 247003 (2010).

\bibitem{Scheer1997}
E. Scheer, P. Joyez, D. Esteve, C. Urbina, and M. H. Devoret,
Phys. Rev. Lett. {\bf 78}, 3535 (1997).

\bibitem{Cron2001}
R. Cron, M. F. Goffman, D. Esteve, and C. Urbina,
Phys. Rev. Lett. 86, {\bf 4104} (2001).

\bibitem{note-exp}
In Ref.~\onlinecite{Chauvin2006} the experimental study focused on the interplay
between microwaves and finite-bias transport (Shapiro steps and subharmonic gap
structure). The effect of the microwaves on the supercurrent was also measured,
but it was not analyzed in detail.

\bibitem{Chauvin2007}
M. Chauvin, P. vom Stein, D. Esteve, C. Urbina, J. C. Cuevas, and A. Levy Yeyati,
Phys. Rev. Lett. {\bf 99}, 067008 (2007).

\bibitem{Cuevas2002}
J. C. Cuevas, J. Heurich, A. Martin-Rodero, A. Levy Yeyati and G. Sch\"on,
Phys. Rev. Lett. {\bf 88}, 157001 (2002).

\end{thebibliography}
\end{document}